\documentclass{article}
\pdfoutput=1
\usepackage[top=1in, bottom=1in,left=1in,right=1in]{geometry}
\usepackage[english]{babel}
\usepackage[utf8]{inputenc}
\usepackage{amsmath}
\usepackage{amsbsy}
\usepackage{algpseudocode}
\usepackage{float}
\usepackage{natbib}
\usepackage{dirtytalk}
\usepackage{mathrsfs}
\usepackage{epigraph}
\usepackage{listings}
\usepackage[export]{adjustbox}
\usepackage{hyperref}
\hypersetup{
    colorlinks=true,
    linkcolor=blue,
    filecolor=magenta,      
    urlcolor=green,
    citecolor=brown,
    pdftitle={Making heads or tails of systemic risk measures},
    pdfauthor={Aleksy Leeuwenkamp},
   pdfstartpage=1
    }
\usepackage[nottoc,numbib]{tocbibind}
\usepackage{amssymb}
\usepackage{amsthm}
\usepackage{bbm}
\usepackage{verbatim}
\usepackage[mathscr]{euscript}
\usepackage{graphicx}
\usepackage{censor}
\usepackage[colorinlistoftodos]{todonotes}
\usepackage{url}
\usepackage{subfigure}

\numberwithin{myth}{subsection}
\newtheorem{myprop}{Proposition}
\numberwithin{myprop}{subsection}

\numberwithin{mylemma}{subsection}
\theoremstyle{definition}
\newtheorem{mydef}{Definition}
\numberwithin{mydef}{subsection}

\numberwithin{mymod}{subsection}

\numberwithin{myrem}{subsection}

\numberwithin{myex}{subsection}

\DeclareMathOperator{\var}{Var}
\DeclareMathOperator{\Expect}{\mathbb{E}}

\DeclareMathOperator{\Real}{\mathbb{R}}
\DeclareMathOperator{\CoV}{CoVaR}
\DeclareMathOperator{\DCov}{\Delta\text{-CoVaR}}
\DeclareMathOperator{\VaR}{VaR}
\DeclareMathOperator{\CoES}{CoES}
\DeclareMathOperator{\DCoES}{\Delta\text{-CoES}}
\DeclareMathOperator{\ES}{ES}
\DeclareMathOperator{\MES}{MES}
\providecommand{\keywords}[1]{\textbf{\textit{Keywords---}} #1}
\providecommand{\classificationjel}[1]{\textbf{\textit{JEL---}} #1}
\title{Making heads or tails of systemic risk measures}
\author{Aleksy Leeuwenkamp\footnote{ KU Leuven Faculty of Economics and Business (FEB) Department of Accounting, Finance \& Insurance (AFI),\newline Naamsestraat 69 3000 Leuven Belgium  Email: aleksy.leeuwenkamp@kuleuven.be . An earlier version of this paper has been previously circulated under the title "$\DCoES$". The author thanks Stijn Claessens, Lars Van Cutsem, Roman Gonachrenko , Ralph De Haas, Hamza Hanbali, Florian Hoffmann,Wentao Hu, C\'{e}dric Huylebroek, Sotirios Kokas, Ivan Markovi\'{c}, participants of the BSE Banking Summer School 2022, the CEBA Seminar at the University of St. Petersburg the IFABS 2022 conference in Naples and the 17th Belgian Financial Research Forum 2023 for comments, suggestions and other forms of assistance.  }  }
\date{April 2023}

\begin{document}

\maketitle
\begin{abstract}
   \noindent This paper shows that the $\CoV,\DCov,\CoES,\DCoES$ and $\MES$ systemic risk measures can be represented in terms of the univariate risk measure  evaluated at a quantile determined by the copula. The result is applied to derive empirically relevant properties of these measures concerning their sensitivity to power-law tails, outliers and their properties under aggregation. Furthermore, a novel empirical estimator for the $\CoES$ is proposed. The power-law result is applied to derive a novel empirical estimator for the power-law coefficient which depends on $\DCoES/\DCov$. To show empirical performance simulations and an application of the methods to a large dataset of financial institutions are used. This paper finds that the $\MES$ is not suitable for measuring extreme risks. Also, the $\ES$-based measures are more sensitive to power-law tails and large losses. This makes these measures more useful for measuring network risk but less so for systemic risk. The robustness analysis also shows that all $\Delta$ measures can underestimate due to the occurrence of intermediate losses. Lastly, it is found that the power-law tail coefficient estimator can be used as an early-warning indicator of systemic risk.   \newline
   
\end{abstract}
\newline
\classificationjel{C21, C51, C58, E32, G01, G12, G17, G20, G32}\newpage
\section{Introduction}\label{SEC:intro}
Accurately and reliably measuring systemic risk is still a surmountable challenge after the Great Financial Crisis (GFC) in 2008. The main lesson from the GFC was that in order to have an accurate assessment of systemic risk it is insufficient to measure and control the risk of individual financial institutions (microprudential). Considering the risk in the system as a whole (macroprudential) has become a vital strand in the literature as can be seen in \cite{Macpruclaessens} and \cite{sysriskbook} for example. As a consequence, over the years following the GFC the attention has shifted to attempting to measure systemic risk and more precisely risk emerging from the interactions between financial institutions and the networks these interactions create.\\

\noindent Two very popular methods that have emerged in the literature are the CoVaR method proposed by \cite{AdrianBrunnCovar} and the MES method proposed by \cite{MES1}. Both of these systemic risk measures are conditional in the sense that rather looking at the risk of a financial institution $Y$ in isolation they consider the risk of $Y$ conditional on another institution $X$ or the financial system being in a state of distress. While both of these papers provided solid economic foundations for their respective measures the mathematical and statistical properties were not elaborated much upon. Despite this, both measures have become a stable of the empirical literature on systemic risk and even beyond \footnote{ For the CoVaR applications some recent examples: \cite{ji2018uncertainties,brunnermeier2020asset,beck2020bank,NetwNN,TORRI2021104125,zelenyuk2022effects,song2022temperature}. The MES has since its inception been implemented in the SRISK systemic risk measure \cite{MES2}  and in the systemic risk suite V-lab \cite{vlab}. }. As a consequence, a considerable amount of literature is devoted to this topic \footnote{ Some references: \cite{GIRARDI20133169,mainik2014dependence,BERNARD2015104,BERNARDI20178,Jaworski+2017+1+19,SORDO2018105,dhaene2022systemic}}. However, as far as the author is aware results from this literature have only been sparsely applied in the empirical literature such as \cite{REBOREDO20132665,REBOREDO2015214,CopulaGarchCovar}. Also, as both systemic risk measures aim to measure (conditional) tail risks and tail dependencies in (sometimes aggregated) financial data some obvious but unanswered questions are: "How good are these measures at actually capturing these risks?", "What effects do very large losses in the data have on the estimates?", "How does aggregation affect the risk measures?" and "What are the practical implications of these questions?". Answering these questions is crucial in order to assess the empirical usefulness and reliability of these risk measures. In the case of the unconditional $\VaR$ and $\ES$ these questions have been answered already in \cite{yamai2002comparative,mcneil2015quantitative}, \cite{ContRobrisk} , \cite{Cohriskmeas,embrechts2009multivariate,mcneil2015quantitative} and \cite{Danielsson2001,DANIELSSON2008321} respectively. However, for conditional risk measures like the $\CoV,\CoES$ and $\MES$ the questions remain open. The answers to these questions form part of the reason behind the regulatory push from $\VaR$ to $\ES$-based risk measures for unconditional risks \cite{FTRB}. \\

\noindent By applying copula (\cite{nelsen2007introduction}), extreme value (\cite{ModexteventsEmbr}) and statistical robustness theory (\cite{hampel1971general})  this paper aims to, on one hand, contribute to the literature general mathematical, extreme value properties and robustness of risk measures \footnote{ \cite{Cohriskmeas,yamai2002comparative,embrechts2009multivariate,ContRobrisk,mainik2014dependence,BERNARD2015104,BERNARDI20178,Jaworski+2017+1+19,SORDO2018105,dhaene2022systemic}}. On the other hand, by showing the practical implications of the results and proposing new estimators  this paper aims to contribute to the literature on model risk and its economic implications \footnote{\cite{Danielsson2001,DANIELSSON2008321,embrechts_2010}}, conditional risk measure estimation \footnote{\cite{GIRARDI20133169,AdrianBrunnCovar,MES1,CopulaGarchCovar,TORRI2021104125,NetwNN})}, (dynamic) power-law coefficient estimation \footnote{ \cite{hill1975simple,kelly2014dynamic}} and the development of systemic risk measures and early-warning indicators \footnote{ For comprehensive surveys see:\cite{bisias2012survey,benoit2017risks}}. The final goal is to offer empiricists a simple but sound set of facts and results to aid their choice, estimation and interpretation of conditional risk measures while also pointing out some shortfalls. \\

\noindent The paper is structured as follows: in section \ref{Sec:covcoes} the mathematical theory and the results are introduced and proven. In section \ref{Sec:methods} the estimators are proposed and a Monte-Carlo simulation setups to test their properties and the theoretical results are provided. In section \ref{Sec:data} the data are discussed. In section \ref{Sec:results} all the empirical results are provided and discussed. In section \ref{Sec:conc} a short conclusion is provided. 
\section{Theory: the CoVaR and CoES}\label{Sec:covcoes}
\subsection{Definition}\label{Ssec:def}
In this paper will make use of the profit/loss P/L approach like in \cite{AdrianBrunnCovar}. This means that if $P$ is a random variable representing returns or payoffs then $X=-P$ represents losses. Then the Value-at-Risk ($\VaR$) of the losses at a significance level $\alpha\in(0,1)$ can be represented as follows:
\begin{align*}
    \VaR_\alpha(X)=F^{-1}_X(\alpha)
\end{align*}
\noindent As in this setting the interest lies in high quantiles far above $0.5$ such as $\{0.95,0.975,0.99\}$ the $\VaR$ will always be positive. In \cite{AdrianBrunnCovar} the $\CoV$ is proposed in order to capture the effect of financial institutions affecting the financial system or another institution. Simply, the $\CoV$ at a level $\beta\in(0,1)$ is the $\VaR$ of the financial system or another institution $Y$ given that an institution $X$ is in distress. \cite{AdrianBrunnCovar} propose to use the condition $X=\VaR_\alpha(X)$ to denote $X$  being in distress. Hence, their definition of $\CoV$ is
\begin{align*}
    \CoV^=_{\alpha,\beta}(Y\mid X)= \VaR_\beta(Y\mid X=\VaR_\alpha(X)).
\end{align*}
\noindent While this conditioning makes estimation of the $\CoV$ easy as quantile regression can be used it has a set of problems which were first highlighted in \cite{mainik2014dependence}. First, conditioning on a set of measure zero makes the measure more involved in a probabilistic sense and less stable in a statistical sense as discussed in \cite{GIRARDI20133169} and \cite{mainik2014dependence}. Second, this version of the measure even in the most simple case of bivariate normal returns is not guaranteed to increase as the underlying random variables become more dependent i.e. it is not dependence consistent. This property makes this version of the measure unreliable in any empirical application and also risky to use in regulation as financial institutions could lower their $\CoV$ by becoming more dependent with other institutions. In order to partly fix this issue and to establish a measure that measures an institutions risk contribution to that of another or the system \cite{AdrianBrunnCovar} propose the $\DCov$:
\begin{align*}
    \DCov^\text{=,med}_{\alpha,\beta}(Y\mid X)= \CoV^=_{\alpha,\beta}(Y\mid X)-\CoV^=_{1/2,\beta}(Y\mid X).
\end{align*}
\noindent In simpler terms, this is the change in the losses of $Y$ that can occur with probability $1-\beta$ given that the losses of $X$ move from their median level to their $\VaR_\alpha$ level. \cite{AdrianBrunnCovar} show that in the case of bivariate normal returns this measure is increasing in the correlation coefficient $\rho$. However, \cite{mainik2014dependence} argue that this result is superficial as it is proportional to the traditional CAPM beta. Furthermore, they show that once one deviates from this simple model the dependence consistency result does not hold and propose different definitions of the $\CoV$ and $\DCov$ which are dependence consistent under a broader class of distributions and also remediate the probabilistic and statistical issues by simply changing the conditioning from $X=\VaR_\alpha(X)$ to $X\geq \VaR_\alpha(X)$. Hence, they obtain:
\begin{equation}\label{eq:defcov}
     \CoV_{\alpha,\beta}(Y\mid X)= \VaR_\beta(Y\mid X\geq \VaR_\alpha(X))
\end{equation}
and 
\begin{equation*}
     \DCov^\text{med}_{\alpha,\beta}(Y\mid X)= \CoV_{\alpha,\beta}(Y\mid X)-\CoV_{1/2,\beta}(Y\mid X).
\end{equation*}
Hence, in the rest of the paper the definition of the $\CoV$ in equation \ref{eq:defcov} will be used. In \cite{GIRARDI20133169} it is shown that these measures can be estimated using bivariate GARCH models. Even with this change the $\DCov$ is not ideal yet as it is only dependence consistent under still quite strict assumptions (see \cite{SORDO2018105,dhaene2022systemic}), the second term opens up a conundrum of what the appropriate benchmark state for $X$ is (see:\cite{GIRARDI20133169}) and the interpretation of the risk contribution is not really statistical in nature. In the literature another version of the $\DCov$ has been proposed
\begin{equation}\label{eq:dcovdef}
    \DCov_{\alpha,\beta}(Y\mid X)= \CoV_{\alpha,\beta}(Y\mid X)-\VaR_\beta(Y).
\end{equation}
In \cite{SORDO2018105,dhaene2022systemic} it is proven that this version is dependence consistent under more general assumptions on the multivariate distribution. Because the second term now provides a baseline assuming independence of $X$ and $Y$ the $\DCov$ has a natural statistical interpretation and it makes the choice of benchmark state for $X$ irrelevant. Now, the $\DCov$ can be interpreted as the difference in the losses of $Y$ which occur with probability of $1-\beta$ if $X$ is in distress and $X$ and $Y$ are dependent versus if $X$ is in any state and $X$ and $Y$ are independent. This implies that the measure really shows the amount of risk that $X$ and its dependence with $Y$ pose on $Y$ (total risk) minus the risk of $Y$ itself (microprudential). Hence, this interpretation fits more with the spirit of macroprudential financial regulation which was the original intended purpose of the $\DCov$ \cite{AdrianBrunnCovar}. Therefore, in the rest of the paper we will use the definition in equation \ref{eq:dcovdef} for the $\DCov$. \\

\noindent Analogously to the univariate case there also exist co-risk versions of the Expected Shortfall ($\ES$) introduced in \cite{acerbiES1}. These are the $\CoES$ and $\DCoES$:
\begin{equation}\label{eq:defcoes}
     \CoES_{\alpha,\beta}(Y\mid X)= \Expect[Y\geq \CoV_{\alpha,\beta}(Y\mid X)\mid X\geq \VaR_\alpha(X) ]=\frac{1}{1-\beta}\int_{\beta}^1 \CoV_{\alpha,q}(Y\mid X) dq,
\end{equation}
\begin{equation}\label{eq:defdcoes}
        \DCoES_{\alpha,\beta}(Y\mid X)= \CoES_{\alpha,\beta}(Y\mid X)-\ES_\beta(Y).
\end{equation}
\noindent The interpretation of the $\CoES$ is the expected loss that occurs with probability $1-\beta$ if $X$ and $Y$ are dependent and $X$ is in distress. The $\DCoES$ then computes the difference of this expected loss with the expected loss if $X$ and $Y$ were independent. In this paper these definitions for both will be used. In the next section an alternative representation of the $\CoES$ will be proposed which greatly reduces the complexity of these measures when proving properties and estimation. The $\CoES$ and $\DCoES$ have similar properties to the $\CoV$ and $\DCov$ in that under this conditioning on $X$ both are dependence consistent under very general assumptions on the multivariate distribution \cite{mainik2014dependence,SORDO2018105,dhaene2022systemic}. However, the $\DCoES$ is dependence consistent under more general assumptions on the marginal distribution of $Y$. In empirical work this measure has been used most notably in \cite{CopulaGarchCovar}. However, in the literature no attention has been given yet to the statistical properties of the $\CoV,\DCov,\CoES$ and $\DCoES$.   
\subsection{Representation in terms of the copula}\label{Ssec:cop}
A downside of the expressions of the $\CoV$ and $\CoES$ given in the previous section is that they make the mathematics and statistics unnecessarily complicated. For example in \cite{AdrianBrunnCovar} only under the restrictive assumption of a linear relationship between $X$ and $Y$ and a bivariate normal distribution an explicit could be found for the $\CoV^=$ and $\DCov^{=,med}$. This paper extends their results in the Appendix to the multivariate t-distribution. In \cite{mainik2014dependence} the same was attempted for the $\CoV$ but this resulted in integrals that could not be analytically solved. However, using copulas to model the multivariate distribution \cite{BERNARDI20178} found very simple representations for both the $\CoV^=$ and $\CoV$. \\

\noindent To start off, a copula is an alternative representation of the multivariate distribution function of random variables. In simple terms a copula shows more explicitly how two or more marginal distributions are linked together. Therefore, copulas are natural objects for studying dependence structures and are already popular in finance and other fields \cite{genest2007everything,embrechts2009,Genest2009,mcneil2015quantitative}. A copula $C:[0,1]^d\xrightarrow[]{}[0,1]$ is a function from $d$ uniform margins to $[0,1]$. When $d=2$ it is related to the distribution function as follows \cite{sklar1959fonctions}:
\begin{align*}
    \mathbb{P}(X\leq x,Y\leq y)=F_{X,Y}(x,y)=C_{X,Y}(U\leq F_X(x), V\leq F_Y(y)).
\end{align*}
From the definition \footnote{This representation is unique if $X$ and $Y$ are continuous.} it can already be seen that a copula and its margins are quite independent in the sense that one can use the same copula but with different $F_X$ and $F_Y$. While this will result in a different multivariate distribution of course the way in which $X$ and $Y$ are linked will be the same. In fact, copulas are invariant to a broad range of transformations (increasing) of the marginals \cite{nelsen2007introduction}. Therefore, the copula represents the entire dependence structure between $X$ and $Y$ \cite{nelsen2007introduction}. Another convenient property of copulas are the Fr\'{e}chet-Hoeffding bounds. These state that every copula is bounded below and above by the two cases of maximal positive and minimal negative dependence. Therefore, for a given copula $C(u,v)$ we know that
\begin{align*}
    W(u,v)\leq C(u,v)\leq M(u,v) \text{ for all }(u,v)\in[0,1]^2.
\end{align*}
Where $W(u,v)=\max(u+v-1,0),M(u,v)=\min(u,v)$ with the lower(upper) bound being the copula of two counter-monotonic(comonotonic) random variables. Two random variables $X$ and $Y$ are said to be counter-monotonic(comonotonic) if there exists a decreasing(increasing) function $f(.)$ such that $Y=f(X)$. For example, if $f$ is linear and $\rho(X,Y)=1$ then $X$ and $Y$ are comonotonic. As $f$ does not have to be linear copulas capture any possible kind of dependence and hence are more general than the bivariate normal distribution\footnote{In fact, if the copula is Gaussian and the margins are normal then the distribution is equivalent to a bivariate normal distribution.}. This paper will focus on a positive dependence structure so for any $C(u,v)$ it holds that $I(u,v)\leq C(u,v)\leq M(u,v)$ with $I(u,v)=uv$ the independence copula.  Now the result of \cite{BERNARDI20178} adapted to the P/L setting is as follows:
\begin{equation*}
   1-F_{Y\mid X\geq\VaR_\alpha(X)}(\CoV_{\alpha,\beta}(Y\mid X))= \mathbb{P}(Y\geq \CoV_{\alpha,\beta}(Y\mid X)\mid X\geq \VaR_{\alpha}(X))=\frac{\Bar{C}(\alpha,F_y(\CoV_{\alpha,\beta}(Y\mid X)))}{1-\alpha}=1-\beta.
\end{equation*}
Let $F_y(\CoV_{\alpha,\beta}(Y\mid X))=\omega$ with $\omega(\alpha,\beta,C)$ the largest solution to the equation $\Bar{C}(\alpha,\omega)=(1-\alpha)(1-\beta)$ with $\Bar{C}(u,v)=1-u-v+C(u,v)$. Then it follows that: 
\begin{equation}\label{def:covrep}
    \CoV_{\alpha,\beta}(Y\mid X)=F^{-1}_Y(\omega)=\VaR_\omega(Y)
\end{equation}
and
\begin{equation}\label{def:dcovrep}
    \DCov_{\alpha,\beta}(Y\mid X)=\VaR_\omega(Y)-\VaR_\beta(Y).
\end{equation}
 A special case of the copula method to obtain the $\CoV$ was used in \cite{GIRARDI20133169}. This result is so powerful because it reduces the conditional problem to a marginal problem with the dependence structure fully captured by $\omega(\alpha,\beta,C)$. In the copula literature often parameterized copulas are used for theoretical and empirical purposes with the parameters denoting the degree of dependence. As an example the Gumbel copula has one parameter $\theta\in[1,\infty)$ with the lower bound attained in the case of independence and the upper limit attained under comonotonicity. In this case $\omega(\alpha,\beta,\theta)$ and in \cite{BERNARDI20178} it is shown that for lower tail dependence it has an analytical solution.   Applying the results of \cite{BERNARDI20178,Jaworski+2017+1+19} we know that for a copula with a positive dependence structure $\beta\leq\omega\leq \alpha+\beta-\alpha\beta$ with the lower bound attained in the case of independence and the upper bound attained in the case of comonotonicity. With equation \ref{def:covrep} one can prove results for the $\CoV$ and $\DCov$ by using results for the marginal $\VaR$.  In estimation, one can reduce the problem of estimating $\CoV$ to fitting a copula, computing $\omega$ and then estimating the $\VaR$ at level $\omega$. The former has a large literature \cite{hofert2019elements} and the fact that a lot of popular copulas have analytical expressions for $\omega$ \footnote{Contrary to the Gaussian and t-copulas, see \cite{GIRARDI20133169}.} while for the latter there exists a rich literature on univariate $\VaR$ estimation methods with known statistical properties \cite{Condvarcompari}. The addition in this paper is to now extend this result to the $\CoES$ which results in the following expression:
\begin{equation}\label{def:coesrep}
    \CoES_{\alpha,\beta}(Y\mid X)=\ES_\omega(Y)=\frac{1}{1-\omega}\int_{\omega}^1\VaR_q(Y) dq
\end{equation}
and
\begin{equation}\label{def:dcoesrep}
     \DCoES_{\alpha,\beta}(Y\mid X)=\ES_\omega(Y)-\ES_\beta(Y).
\end{equation}
Equation \ref{def:coesrep} will be proven in the Appendix section \ref{App:proofESMESrep}. The representations of the risk measures given in equations \ref{def:covrep},\ref{def:dcovrep},\ref{def:coesrep} and \ref{def:dcoesrep} will form the basis for all following results on the statistical properties. Next, we also provide a new representation of the MES by \cite{MES1} which are the latter terms of the equality:
\begin{equation*}
    \MES_\alpha(Y\mid X)=\CoES_{\alpha,0}(Y\mid X)=\ES_{\omega(\alpha,0,C)}(Y)=\frac{1}{1-\omega(\alpha,0,C))}\int_{\omega(\alpha,0,C)}^1\VaR_q(Y) dq
\end{equation*}
\noindent This representation follows that from the $\CoES$ representation and is proven in the Appendix section \ref{App:proofESMESrep}. The proof also shows that the $\MES$ is a special case of the $\CoES$ with $\beta=0$. Hence, unless specifically mentioned in the rest of the paper any property of the $\DCoES$ can be assumed to hold for the $\MES$ as well. 
\subsection{Coherence, (in)dependence, symmetry and invariance}\label{Ssec:mathprop}
Following the representation results in the previous section in this section some additional mathematical properties of the risk measures are proven. The $\VaR$ can be shown to not be coherent in the sense of \cite{Cohriskmeas} due to it not satisfying sub-additvity in general. However, under the condition of a linear combination of elliptically distributed random variables the $\VaR$ is sub-additive \cite{mcneil2015quantitative}. Beyond elliptical distributions in \cite{embrechts2009multivariate} it is shown that establishing (asymptotic) sub-additivity of the $\VaR$ is a complex affair but mostly depends on the heaviness of the tails and the existence of the mean outside the case of elliptical distributions. The $\ES$ is sub-additive and hence coherent in general.  Using the representation of the $\VaR$ it is straightforward to see that the $\CoV$ in general is not coherent. Similarly, using the representation of the $\CoES$  it is straightforward to establish coherence as it inherits the property from the $\ES$. In empirical simulations by \cite{danielsson2005subadditivity} it has been found that the $\VaR$ often does satisfy sub-additivity. This result is strengthened in \cite{danielsson2013fat} where asymptotic subadditivity for regularly varying tails with tail exponent $\xi<1$ is satisfied\footnote{ See section \ref{Ssec:tails} for a definition.}.  Moreover, in \cite{VaRsubaddprob} it is argued that sub-additivity is not a useful or even desirable property in some risk management problems. Hence, in the rest of the paper this property will not not be directly used to inform the choice of risk measure.\\

\noindent Since the $\DCov$ and $\DCoES$ capture the dependence structure as well the next results are about their behavior under independence and comonotonicity. 
\begin{myprop}
Let $(X,Y)$ be a bivariate random vector with copula $C(u,v)$. It will hold that:
\begin{equation*}
    \DCov_{\alpha,\beta}(Y\mid X)=0 \iff X,Y \text{ are independent,}
\end{equation*}
\begin{equation*}
    \DCoES_{\alpha,\beta}(Y\mid X)=0 \iff X,Y \text{ are independent,}
\end{equation*}
\begin{equation*}
    \DCov_{\alpha,\beta}(Y\mid X)\text{ is maximal}\iff X,Y \text{ are comonotonic and } 
\end{equation*}
\begin{equation*}
    \DCoES_{\alpha,\beta}(Y\mid X)\text{ is maximal}\iff X,Y \text{ are comonotonic. } 
\end{equation*}
\end{myprop} 
\noindent The proof is provided in the Appendix. The practical implications are that both risk measures capture the full dependence structure and hence can provide more information than simple correlations. A corollary to this result is that under a positive dependence structure $\DCov$ and $\DCoES$ are bounded below by zero and bounded above by the value under a comonotonic bivariate random vector. The fact that both $\DCoES$ and the $\DCov$ are bounded above by the comonotonic scenario makes developing a market fragility measure based on them and identifying strongly dependent institutions straightforward.\\

\noindent Another property of some dependence measures is that they are symmetric. For example, for the Pearson correlation $\rho(X,Y)$ we always have that $\rho(X,Y)=\rho(Y,X)$. Because the co-risk measures and their risk contribution counterparts also take into account the risk of the variable that is not conditioned on we generally expect these measures to not be symmetric as the distribution of losses of a given institution $Y$ can be very different from that of $X$. The following proposition establishes under what conditions these risk measures are symmetric.
\begin{myprop}
Let $(X,Y)$ be a bivariate random vector. If $C(u,v)=C(v,u)$ for all $(u,v)\in[0,1]^2$ and $F_X=F_Y$ then we have that:
\begin{equation*}
    \CoV_{\alpha,\beta}(Y\mid X)=\CoV_{\alpha,\beta}(X\mid Y), \CoES_{\alpha,\beta}(Y\mid X)=\CoES_{\alpha,\beta}(X\mid Y),
\end{equation*}
\begin{equation*}
    \DCov_{\alpha,\beta}(Y\mid X)=\DCov_{\alpha,\beta}(X\mid Y)\text{ and } \DCoES_{\alpha,\beta}(Y\mid X)=\DCoES_{\alpha,\beta}(X\mid Y).
\end{equation*}

\end{myprop}
\noindent \begin{proof}
Due to the copula $\omega$ will not change if the conditioning is flipped because $\Bar{C}(u,v)=\Bar{C}(v,u)$ for all $(u,v)\in[0,1]^1$ and due to $F_X=F_Y$ the $\VaR_\omega(Y)=\VaR_\omega(X)$ and $\ES_\omega(Y)=\ES_\omega(X)$. This then implies equality of the $\DCov$ and $\DCoES$ as well. 
\end{proof} 
\noindent Random vectors that satisfy copula symmetry and equality of marginal distributions are called exchangeable. This result is important because in the empirical literature often symmetric copulas are used. Therefore, if one also obtains or imposes similar margins the order of $X$ and $Y$ should not matter. Researchers using these risk measures in a network setting should therefore be wary but could also exploit this property to reduce the computational burden of estimation as under symmetry for a set of $n$ institutions only $n(n-1)/2$ $\DCov$'s or $\DCoES$'s have to be estimated.\\    

\noindent Knowing when the risk measures are invariant if $X$ or $Y$ are transformed is crucial as for example the Pearson correlation is only invariant under increasing linear transformations which can lead to surprising results when data are transformed with a non-linear function, see \cite{embrechts2001correlation}. Due to the properties of copulas the resulting multivariate distribution is invariant under increasing transformations on both variables \cite{nelsen2007introduction}. However, the risk measures depend on the distribution of $Y$ so therefore this invariance result only applies to the variable that is conditioned on; $X$. Still this property while more restrictive than the property of copulas is valuable for empirical researchers.  
\subsection{Tail sensitivity}\label{Ssec:tails}
It is well-known that the $\ES$ is more sensitive to the tails of a distribution than the $\VaR$. Alongside coherence it is one of the main reasons the $\ES$ was originally developed, see \cite{acerbiES1} and \citet{ACERBIES2}. In order to assess this difference in tail sensitivity beyond some high quantile extreme value theory will be applied. First, results will be derived when assuming only the tails of the distribution of $Y$ matter. However as the $\CoV$ and $\CoES$ are by definition multivariate we will also consider the case where the tails of both distributions matter.\\

\noindent In the univariate case as the quantiles the risk measures are computed at can be very high indeed (between $\beta$ and $\alpha+\beta-\alpha\beta$) to establish results on tail sensitivity extreme value theory (EVT) is needed. Applying EVT to risk management was proposed in \cite{embrechts1999extreme} and is now widely accepted and developed \cite{embrechts2013modelling,nolde2021extreme}. In extreme value theory there exists a very elegant and robust result regarding the behavior of the tails of a given distribution. This result is called the Theorem of Pickands-Balkema-DeHaan \cite{pickands1975statistical,balkdehaan} and in simple terms it states that for a sufficiently high $u$ one obtains that:
\begin{equation*}
    F_u(y)=\mathbb{P}(Y-u\leq y\mid Y>u)=\begin{cases}1-\left(1+\frac{\xi(y-u)}{s}\right)^{-1/\xi} & \text{ if }\xi\neq 0\\
    1-\exp\left(-\frac{y-u}{s}\right) &\text{ if }\xi=0
    \end{cases}
\end{equation*}

\noindent which is a Generalized Pareto distribution (GPD)\footnote{ Special cases of the GPD are the Pareto distribution for $u=x_m=s/\xi$ and $\alpha=1/\xi$ ,the continuous uniform distribution on $(0,s)$ for $\xi=-1$ and the exponential distribution for $\xi=0$ and $u=0$.} with a tail parameter (index) $\xi$ and scale parameter $s$. The theorem is exact when $u\xrightarrow[]{}\infty$ but an approximation for finite $u$. This result is a special case of a result established by \cite{gnedenko1943distribution}. The result states that for a large class of distributions the tail function behaves as follows for $y\xrightarrow[]{}\infty$: $\bar{F}(Y)=1-F(y)=L(y)y^{-1/\xi}$ for $\xi>0$ with $L(y)$ a function that slowly varies with $y$\footnote{A slowly-varying function satisfies that $L(tx)/L(x)\xrightarrow[]{}1$ for $t\xrightarrow[]{}\infty$.}. The tail index $\xi$ fully determines the heaviness of the tails and distributions that have this tail representation are called regularly varying. For some $\xi>0$  the highest finite moment of $Y$ is $1/\xi$ . Hence, if $\xi>0$ the distribution has a heavy tail that behaves like a power-law whereas if $\xi\leq 0$ the tail either decays at an exponential rate (light-tailed distributions like the normal) or does not exist because the distribution has bounded support. In stock return data the power-law tail is a well-established empirical result with theoretical foundations \footnote{Some references on this are: \cite{PL5,PL6,ContFinanceempirics,PLth,Pl1,PL2,PL4,PL3,PL7} }. In this literature estimates of $\xi\approx 1/3$\footnote{The power-law estimates in these papers are usually in the form of $1/\xi$ so are in the interval $[1,\infty)$.} are obtained which do not seem to depend on the time period or stock market studied. The power-law does change when considering different time intervals with tails converging to Gaussian tails as the interval increases \footnote{See \cite{PL5,PL6,ContFinanceempirics,PL3}.}. Power-law tails of economic variables are common and speculated to be universal in economics in \cite{PL7}. One must mind that these results apply to the \textit{unconditional} returns distribution. More recently, the \textit{conditional} distribution having time-dependent power-law tails has been explored empirically in \cite{kelly2014dynamic,kelly2014tail}\footnote{ For theoretical literature on asset pricing with (varying) tail risk see for example \cite{barro2006rare,gabaix2012variable}.}. Considering the power-law distributions used in the empirical literature are a special case of the GPD it seems to be applicable to model the tails of asset prices both in unconditional and conditional setting with the GPD. The following proposition establishes the properties of the co-risk and risk contribution measures under a GPD tail of $Y$.
\begin{myprop}\label{prop:tailsensuni}
Let $(X,Y)$ be a bivariate random vector with marginals $F_X$, $F_Y$ and a positive dependence structure. Suppose that the tails of $Y$ beyond some point $u=\VaR_\gamma(Y)$ with $\gamma\leq\beta,\gamma>0.5$ follows a GPD with $\mu=u,s>0,\xi\in[0,\infty)$  then 
\begin{equation*}
    \CoES_{\alpha,\beta}(Y \mid X)-\CoV_{\alpha,\beta}(Y \mid X)=s\left( \frac{\left(\frac{1-\omega}{1-\gamma}\right)^{-\xi}}{1-\xi}\right)\geq 0 \text{ if }0<\xi<1
\end{equation*}
\begin{equation*}
     \CoES_{\alpha,\beta}(Y \mid X)-\CoV_{\alpha,\beta}(Y \mid X)=s \text{ if }\xi=0 \text{ or if } X,Y \text{ are independent}
\end{equation*}
\begin{equation*}
     \CoES_{\alpha,\beta}(Y \mid X)-\CoV_{\alpha,\beta}(Y \mid X)=\infty \text{ if }\xi\geq1
\end{equation*}
\begin{equation*}
    \DCoES_{\alpha,\beta}(Y \mid X)-\DCov_{\alpha,\beta}(Y \mid X)=s\left( \frac{\left(\frac{1-\omega}{1-\gamma}\right)^{-\xi}-\left(\frac{1-\beta}{1-\gamma}\right)^{-\xi}}{1-\xi}\right)\geq 0 \text{ if }0<\xi<1
\end{equation*}
\begin{equation*}
     \DCoES_{\alpha,\beta}(Y \mid X)-\DCov_{\alpha,\beta}(Y \mid X)=0 \text{ if }\xi=0 \text{ or if } X,Y \text{ are independent}
\end{equation*}
\begin{equation*}
     \DCoES_{\alpha,\beta}(Y \mid X)-\DCov_{\alpha,\beta}(Y \mid X)=\infty \text{ if }\xi\geq1
\end{equation*}
\begin{equation*}
    \frac{\DCoES_{\alpha,\beta}(Y \mid X)}{\DCov_{\alpha,\beta}(Y \mid X)}=\frac{1}{1-\xi}\text{ if }0\leq \xi<1 \text{ and }\omega>\beta
\end{equation*}
\end{myprop}
\noindent The proof can be found in the Appendix section \ref{App:prooftailsens}. This proposition makes immediately clear that the $\CoES$ and $\DCoES$ are more sensitive to the tails of the distribution of $Y$ and hence more suitable if one assumes the quantile is sufficiently high  for the Pickands-Balkema-DeHaan result to be accurate. The most important result is the ratio between the risk contribution measure as this exclusively depends on $\xi$. Therefore, one can use this ratio to estimate $\xi$ and it indicates to what extent the $\ES$ based measures capture the tail better than the $\VaR$-based measures . For the $\CoES$ and $\CoV$ a similar result exists but it holds only in the limit of $\omega\xrightarrow{}1$. Furthermore, if one uses a t-distribution for the distribution of $Y$ in \cite{Garchevt} it is shown that the tails behave like $cy^{-\nu}$ with $c$ a normalization constant and with $\nu$ the degrees of freedom of the t-distribution. Hence, the tails behave like a GPD with tail index $\xi=1/\nu$. These results could also apply to a conditional GPD with varying $\xi$ where the risk measures are computed at time $t$ given the GPD at time $t$. Because the $\MES$ cannot satisfy the assumption that $\beta\geq \gamma>0.5$ this result does not hold for it. This makes sense as the $\MES$ is just the expected value of the entire distribution conditional on $X\geq\VaR_\alpha(X)$. Only in the case where the random variables are comonotonic (or close to it) we get $\omega(\alpha,0,C)=\alpha$ and this quantile of the distribution of $Y$ could be sufficiently high to be in the region of the GPD tail. This shows that if one requires a measure that is sensitive to the tail shape the $\MES$ is not suitable as only in the extreme case of $\xi\geq1$ it will not be finite anymore as then $\Expect[Y\mid X]$ will not exist. However, in finance the infinite mean scenario does not seem to be applicable due to the estimate of $\xi\approx 1/3$ obtained in the literature. Next, this result will be generalized to bivariate exceedances which will show that the $\MES$ is not suitable for measuring  tail dependence as well.\\

\noindent In the bivariate case one can still use the Pickands-Balkema-DeHaan result for the marginal distributions but now one must also ask what will the distribution of \textit{joint} exceedances converge as the thresholds $u_1,u_2\xrightarrow[]{}\infty$? In other words, what copula will the bivariate exceedances have? Results on this were first established by \cite{EVcop1} and \cite{EVcop2} and showed that there exists a whole family of extreme-value copulas that characterises the dependence structure of bivariate extremes. For more information on extreme-value copulas the reader is referred to \cite{EVcopbook}. A good parametric extreme-value copula for bivariate extremes is the Gumbel (also called logistic) copula and is studied in \cite{Evcopesti1} and \cite{Evcopestitest}. A convenient property of the Gumbel copula is that it ranges between the independence and comonotonic copula. This model has been used in finance as well \cite{EVcopfinapp}. As this extreme-value copula is simple and ranges from independence to comonotonicity it will be used in the next result. 
\begin{myprop}\label{prop:tailsensbiv}
Let $(X,Y)$ be a bivariate random vector with marginals $F_X$, $F_Y$ and Gumbel copula with dependence parameter $\theta\in[1,\infty)$. Suppose that the tails of $Y$ and $X$ beyond some point $(u_1,u_2)=(\VaR_{\gamma_1}(X),\VaR_{\gamma_2}(Y))$ with $\gamma_1\leq\alpha,\gamma_2\leq\beta$ follow GPDs with $\mu_1=u_1,\mu_2=u_2,s_1>0,s_2>0,(\xi_1,\xi_2)\in[0,\infty)^2$  then the results for the risk measures will be the same as those in proposition \ref{prop:tailsensuni} but with $\omega(\alpha,\beta,C_\theta)$ the largest solution to:
\begin{equation*}
    1-\alpha-\omega+\exp\left(-\left(-\ln(\alpha))^\theta+(-\ln(\omega))^\theta\right)^{1/\theta}\right)=(1-\alpha)(1-\beta).
\end{equation*}
\end{myprop}
\noindent The proof can be found in the Appendix in section \ref{App:prooftailsens}. When $\theta=1$ the solution is explicit $\omega=\beta$ and the $\DCov$ and $\DCoES$ are zero because $X$ and $Y$ are independent. If one takes $\theta\xrightarrow[]{}\infty$ $\omega=\alpha+\beta-\alpha\beta$ and the risk contribution measures are maximal because $X$ and $Y$ are comonotonic. In short, this proposition shows how the tail dependence structure affects the risk measures. This result is less likely to be valid as it also requires that the quantile $\alpha$ of $X$ is far enough in the tail of $X$ for the GPD to be a good approximation of exceedances beyond $X$. As before, the $\MES$ suffers from the same deficiencies also in this setting and hence is not suitable for measuring tail dependence and its effects.  
\subsection{Robustness}\label{Ssec:robust}
Next to the tail sensitivity of risk measures another important aspect is their robustness. Robustness is here defined in the sense of robust statistics and in simple terms it means the sensitivity to outliers. In this paper outliers are not seen as contamination of the data but rather as unexpected large losses that fall outside the distribution of the majority of the data (see \cite{hampel1971general}). Another interpretation is conditional in that the losses represent a shift at $t+1$ to the conditional distribution at $t$ while being agnostic to both loss distributions. The intent is to assess how such losses affect estimates of the risk measures while being agnostic towards the distribution of the majority of the data and the distribution of the outliers.  Mind that this is different from the extreme value paradigm in section \ref{Ssec:tails}. The following theory is based on \cite{ContRobrisk} that first applied robust statistics to risk measures. The main object of interest from this paper and robust statistics that will be used to assess the robustness of a risk measure and its estimator is the sensitivity function.
\begin{mydef}\label{def:sensf}
Let $X$ be a random variable with distribution $F(X)$, let $\Delta_l$ be a dirac delta distribution at $l\in \Real$ and $\rho(X)_{eff}$ a risk estimator in the sense of \cite{ContRobrisk}. Then the influence function $S(l;F)$ is:
\begin{equation*}
    S(l;F)=\underset{\epsilon\xrightarrow[]{}0}{\lim}\frac{\rho_{eff}((1-\epsilon)F(X)+\epsilon\Delta_l)-\rho_{eff}(F(X))}{\epsilon}.
\end{equation*}
\noindent With $\rho(X)_{eff}=\underset{n\xrightarrow{}\infty}{\lim}\hat{\rho}(x_1,\dots,x_n)$ and $\hat{\rho}(x_1,\dots,x_n)$ is an estimate of the risk measure $\rho$ on a sample drawn from $X$ i.i.d.
\end{mydef}
\noindent The sensitivity function can be seen as a derivative in a distributional sense and can be interpreted as follows: how will the risk estimate change if some infinitesimal point masses at $l$ are added to the distribution of $X$? In a more practical sense this corresponds to the addition of data points at $l$ in a large sample. This definition of the sensitivity function is asymptotic in that it assumes the sample is infinite. In \cite{ContRobrisk} a finite-sample version is also provided. However, in this paper the focus remains on the infinite sample version to focus on the effects of outliers in isolation from any kind of sampling error. In practice these results provide a best-case scenario and sampling error will be relevant. In \cite{ContRobrisk} the sensitivity function for the $\VaR$ and the $\ES$ are provided given different estimation methods. In this paper, results concerning the historical estimators will be applied and extended because these are distribution agnostic and because the estimation methodology for the $\CoES$ given in section \ref{Ssec:estim} will be a historical estimator. The robustness results for the $\CoV$ and $\CoES$ will be given by the following proposition.
\begin{myprop}\label{prop:robcoes}
Let $(X,Y)$ be a bivariate random vector with a PD copula $C(u,v)$ and margins $F_X,F_Y$. Then the sensitivity functions $S_1(l),S_2(l)$ for the $\CoV_{\alpha,\beta}(Y\mid X)$ and $\CoES_{\alpha,\beta}(Y\mid X)$ respectively will be:
\begin{equation*}
    S_1(l)=\begin{cases}
        \frac{\omega}{f_Y(\VaR_\omega(Y))} & \text{if } l>\VaR_\omega(Y)\\
        0 & \text{if } l=\VaR_\omega(Y)\\
        \frac{\omega-1}{f_Y(\VaR_\omega(Y))} &\text{ if } l<\VaR_\omega(Y)
        
    \end{cases}
\end{equation*}
\noindent With $f(.)$ the pdf of $Y$ and:
\begin{equation*}
    S_2(l)=\begin{cases}
    \frac{l}{1-\omega}-\frac{\omega}{1-\omega}\VaR_\omega(Y)+\ES_\omega(Y) & \text{ if }l\geq\VaR_\omega(Y)\\
    \VaR_\omega(Y)+\ES_\omega(Y) & \text{ if } l\leq\VaR_\omega(Y)
    
    \end{cases}
\end{equation*}
\end{myprop}
\noindent  The proof can be found in the Appendix section \ref{App:proofrob} but it is straightforward and follows from adapting the results of \cite{ContRobrisk} to the P/L setting and applying the representation results from section \ref{Ssec:cop}. From this proposition it becomes apparent that just as in \cite{ContRobrisk} the $\CoES$ is more sensitive than the $\CoV$ because its sensitivity function is linear beyond $\VaR_\omega(Y)$ in the size of the loss $l$ whereas the sensitivity function of the $\CoV$ is a piecewise constant in $l$. The reason behind this difference is rather simple: the historical $\VaR$ estimator uses the empirical quantile function estimate at a given quantile $\alpha$ whereas the historical $\ES$ takes a sample average of the point exceeding the empirical quantile estimate at a level $\alpha$. Hence, one only needs one data point beyond $\VaR_\alpha$ to make the $\ES$ arbitrarily large whereas with the $\VaR$ estimate one would need more than $(1-\alpha)n$ data points. The result shows that the $\MES$ can react to insignificantly low losses and even to profits as in the case of the $\MES$ $0\leq \omega\leq \alpha$. The sensitivity of the risk contribution measures will be investigated next. The sensitivity function of the $\DCov$ is left out because it will be piecewise constant again and therefore not very interesting. These are given in the next proposition:
\begin{myprop}\label{prop:robdcoes}
Let $(X,Y)$ be a bivariate random vector with copula $C(u,v)$ and margins $F_X,F_Y$. Then the sensitivity function $S_3(l)$ for the $\DCoES_{\alpha,\beta}(Y\mid X)$  will be:
\begin{equation*}
    S_3(l)=\begin{cases}
       \frac{l(\omega-\beta)}{(1-\omega)(1-\beta)}-\frac{\omega}{1-\omega}\VaR_\omega(Y)+\frac{\beta}{1-\beta}\VaR_\beta(Y)+\ES_\omega(Y)-\ES_\beta(Y) & \text{if } l\geq\VaR_\omega(Y)\geq\VaR_\beta(Y)\\
        \VaR_\omega(Y)+\ES_\omega(Y)-\frac{l}{1-\beta}+\frac{\beta}{1-\beta}\VaR_\beta(Y)-\ES_\beta(Y) & \text{if } \VaR_\omega(Y)\geq l\geq \VaR_\beta(Y)\\
       \VaR_\omega(Y)-\VaR_\beta(Y)+\ES_\omega(Y)-\ES_\beta(Y) &\text{ if } \VaR_\omega(Y)\geq\VaR_\beta(Y)\geq l
        
    \end{cases}
\end{equation*}

\end{myprop}
\noindent This proposition is proven in the Appendix section \ref{App:proofrob} but follows straightforwardly from the results in proposition \ref{prop:robcoes} and the definition of the $\DCoES$. In Section \ref{SSec:simresults} the sensitivity of both will be assessed by means of simulations and compared to the theoretical results. The proposition shows that the sensitivity of the $\DCoES$ to large values depends both on the value of the loss $l$ and the dependence structure. Similarly to the $\CoES$ the $\DCoES$ reacts linearly to losses above $\VaR_\omega(Y)$ but less strongly because the second term in the $\DCoES$ dampens this reaction.The dampening and reaction perfectly cancel out when $\omega=\beta$ which is when $S_3(l)=0$ for all $l\in\Real$\footnote{ If $\omega=\beta$ the middle piece of $S_3(l)$ becomes redundant as $l$ can only then be higher or lower than $\VaR_\beta$(Y).}. Also, the strength of the linear reaction is bounded due to the boundedness of $\omega$ and occurs in the comonotonic scenario. In this scenario the length of the middle piece of $S_3(l)$ is also maximized. More notably, in between $\VaR_\omega(Y)$ and $\VaR_\beta(Y)$ the $\DCoES$ reacts linearly in a negative way. This shows the trade-off one must make with the $\DCoES$: on the upside one obtains high sensitivity to large losses which could be useful for early-warning systems but also mind the downside of the increased sensitivity of the $\DCoES$ as for intermediate losses it could react adversely and push estimates downwards. This contrasts with the more stable behavior of the $\DCov$. Interpreting the results in a conditional setting implies that the $\DCov$ will have slower decaying autocorrelations than the $\DCoES$ as even very large losses will not easily affect estimates much unless over time enough of them have occurred to shift the conditional distribution. The $\DCoES$ on the other hand can already react to individual large losses and hence estimates will be less correlated over time. Therefore, combined with the results in section \ref{Ssec:tails} the $\DCov$ seems more reasonable for its intended purpose, measuring long-term systemic risk build-up, while the $\DCoES$ seems to be more fit as a short-term early-warning system.   
\subsection{Aggregation properties}\label{Ssec:aggr}
In the previous sections the properties of the risk measures if $Y$ denotes the losses of a single institution have been studied. However, often in empirical research \footnote{ For example see \cite{AdrianBrunnCovar,CopulaGarchCovar,beck2020bank,brunnermeier2020asset}} $\bar{Y}$ which denotes denotes some weighted average of losses is used. Therefore, it is also necessary to know the behavior of the risk measures under aggregation. First denote $\bar{Y}=\displaystyle\sum_{i=1}^N a_iY_i$ with $ a_i\in[0,1],\displaystyle\sum_{i=1}^N a_i=1$. Using the results in \cite{PL3,PL7} and \cite{jessen2006regularly} it can established that if the tails of each $Y_i$ follow a GPD with tail exponent $\xi_i$ then $\bar{Y}$ has a tail exponent of $\bar{\xi}=\max(\xi_1,\dots,\xi_N)$. Therefore, the tail of $\bar{Y}$ is fully determined by the heaviest tail(s) of the $Y_i$. One can then apply the results from section \ref{Ssec:tails} to obtain the behavior of the risk measures of $\bar{Y}$ given some $X$. Therefore, it can be expected that in a network setting the $\DCoES$ and $\DCov$ will give more diverse results depending on the tail exponent of the $Y$ institution whereas in the system setting both will be very similar given any $X$ and only differ in function of the dependence structure. In any case the $\DCoES$ will be more sensitive to the tails than the $\DCov$. \\

\noindent Concerning robustness under aggregation sub-additivity will become important as this property determines the behavior of the risk measures under aggregation. First, observe that $\DCov_{\alpha,\beta}(\bar{Y}\mid X)\leq \VaR_\omega(\bar{Y})$ and  $\DCoES_{\alpha,\beta}(\bar{Y}\mid X)\leq \ES_\omega(\bar{Y})$. Then, under general conditions it can be deduced \footnote{ Using results of the $\VaR$ and $\ES$ regarding positive homogeneity and sub-additivity, see \cite{Cohriskmeas,mcneil2015quantitative}.} that: $\VaR_\omega(\bar{Y})>\displaystyle\sum_{i=1}^N a_i\VaR_\omega(Y_i)$ and $\ES_\omega(\bar{Y})\leq\displaystyle\sum_{i=1}^N a_i\ES_\omega(Y_i)$. Therefore, it follows that in general the $\DCov$ could exceed the weighted average of $\VaR$s \footnote{As stated in section \ref{Ssec:mathprop} the $\VaR$ can be (asymptotically) sub-additive under some conditions discussed in \cite{embrechts2009multivariate,mcneil2015quantitative}. }. However it also follows that the $\DCoES$ is bounded above by the weighted average of the $\ES$s in general. Suppose that one or more of the $Y_i$s is(/are) perturbed by outlier(s) $l_i$. Then, using the setting and the results of section \ref{Ssec:robust} it can quickly be seen that the $\DCov$ will not react much to the outlier(s) as the bound will be a weighted average of the piecewise constant sensitivity function of the individual $\VaR$s. The weighted average will dampen the effect of any jump in the piecewise constant functions. The dampening is strengthened in the case if conditions hold such that the $\VaR$ is (asymptotically) sub-additive as then the $\DCov$ will be bounded by this weighted average \footnote{ In \cite{embrechts2009multivariate,danielsson2013fat} it is shown that the relatively mild assumption of regularly varying distributions with $\xi<1$ is sufficient to guarantee asymptotic sub-additivity. As in these applications the quantiles can be very high indeed these results can hold approximately. }. However, the  reaction of the $\DCoES$ will be markedly different. The upper bound of the $\DCoES$ will shift upwards as some of the $l_i$'s will be in the region where the individual $\ES$ estimates will react linearly resulting with upper bound being a weighted average of said linear reactions. Again, the weighted average will dampen some of the individual reactions. As seen in proposition \ref{prop:robdcoes} the $\DCoES$ will react more dampened depending on $\omega$ with the possibility of a negative reaction in a certain region. In any case if $l_i$ is sufficiently large the effect will still be linear in $l_i$ but now further dampened by the weighted average. Hence, on aggregated data the $\DCoES$ is still expected to be less robust than the $\DCov$ but both are expected to be more robust than their respective counterparts in the network setting. \\

\noindent Based on these results, it can be stated that the $\DCoES$ will be more suitable for being an early-warning measure of network risk between individual financial institutions than the $\DCov$ due to its higher sensitivity to tails and lower robustness. In the system setting the $\DCoES$ is still more sensitive to the tails and less robust than the $\DCov$ but the differences with the $\DCov$ might be less pronounced. \newpage
\subsection{An extension with varying prudence}
Given the properties of the risk measures outlined in previous sections a practitioner or regulator might want to have some mixture between the properties of the $\VaR$- and $\ES$-based measures with mixture weights that vary over time. This approach could allow for a varying degree of prudence where the higher weight is put on the more sensitive $\ES$-based measures during periods where higher prudence is desirable. Due to the representation results of the $\CoV$ and $\CoES$ such a mixture would boil down to a simple convex combination of the $\VaR$ and $\ES$ at level $\omega$\footnote{ Extending to a convex combination of the risk contribution measures is quite straightforward.}. The resulting risk measure has been proposed in \cite{hu2022tail} and is called the Slide$\VaR$. Hence, in this context it is proposed to name the resulting risk measures the Slide$\CoV$ and Slide$\DCov$.

\section{Methodology}\label{Sec:methods}
\subsection{Estimators}\label{Ssec:estim}
Simply stated, the $\CoES$ estimator proposed in this paper is an extension of the historical $\ES$ estimator from for example \cite{ESestgen}:
\begin{equation*}
    \hat{\ES}_\alpha(Y)=\frac{1}{\lfloor n(1-\alpha)\rfloor}\displaystyle\sum_{i=\lfloor n\alpha\rfloor}^N Y_{(i)},
\end{equation*}
\noindent where $\lfloor v\rfloor$ is the floor function and $Y_{(1)}\leq Y_{(i)}\leq Y_{(N)}$ the $i$-th order statistic. In other words, the historical $\ES$ estimate at a level $\alpha$ is the sample mean of all the observations exceeding the historical $\VaR$ estimate (empirical quantile) at level $\alpha$. Using the representation result the proposed estimator is defined as follows
\begin{equation}\label{eq:CoESest}
    \hat{\CoES}_{\alpha,\beta}(Y\mid X)=\hat{\ES}_{\hat{\omega}}(Y)=\frac{1}{\lfloor n(1-\hat{\omega})\rfloor}\displaystyle\sum_{i=\lfloor n\hat{\omega}\rfloor}^N Y_{(i)}.
\end{equation}
\noindent Mind that in this estimator $\omega$ has to be estimated as well since in practice the true copula and hence $\omega$ are unknown. An estimator of $\DCoES$ is then $\hat{\ES}_{\hat{\omega}}-\hat{\ES}_\beta$. If the data are i.i.d. the historical method is known to have the best statistical performance compared to model-based methods \cite{ESestgen}. Even in the case of dependent observations the historical $\ES$ at least outperforms kernel-based methods \cite{NonparESest} in most scenarios. A downside of the historical $\ES$ is the variance incurred due using a very small amount observations at high quantiles and the bias incurred to its sensitivity to large losses \cite{ContRobrisk}. However, for the purposes of this paper the $\CoES$ estimator must be sensitive to large losses as the $\VaR$ estimators are already robust in this regard. In an unconditional setting obtaining estimates for $\omega$ can be done by estimating the copula of $X$ and $Y$ using the empirical beta copula estimator from \cite{SEGERS201735}. Among the most popular empirical copula estimators this estimator  has the best finite-sample performance in the MSE sense while having the same asymptotic distribution and being smooth \footnote{ Smooth in the sense that for a given sample size $n$ the $n$-th derivative exists. This smoothness also allows for a wider choice of root finding algorithms when solving $\hat{\bar{C}}(\alpha,\omega)=(1-\alpha)(1-\beta)$ in terms of $\omega$. The classical empirical copula estimator is not even continuous while the empirical checkerboard copula estimator is not differentiable everywhere. There also exists the empirical Bernstein copula estimator which has similar smoothness properties to the empirical beta copula estimator. However, in \cite{SEGERS201735} it is shown that in most cases the finite sample performance of the empirical beta copula estimator is better in terms of MSE.   }. Once $\hat{C}(u,v)$ is obtained then $\hat{\omega}$ can be obtained by solving $\hat{\bar{C}}(\alpha,\omega)=(1-\beta)(1-\alpha)$ in terms of $\omega$. The method outlined is a fully nonparametric method and will be applied to obtain the unconditional results. If one requires a parametric copula then these can be fitted using the methods discussed in \cite{hofert2019elements}. Estimation of the unconditional case will be covered in the next section. \\\newpage

\noindent Based on the results in proposition \ref{prop:tailsensuni} also an estimator for $\xi$ and $\xi_t$ can be formulated:
\begin{equation}\label{eq:xiesti}
    \hat{\xi}=\frac{\frac{\hat{\DCoES}_{\alpha,\beta}(Y\mid X)}{\hat{\DCov}_{\alpha,\beta}(Y\mid X)}-1}{\frac{\hat{\DCoES}_{\alpha,\beta}(Y\mid X)}{\hat{\DCov}_{\alpha,\beta}(Y\mid X)}}.
\end{equation}
\noindent This estimator can also estimate $\xi_t$ if time-dependent estimators for the risk measures have been used. If the conditions of proposition \ref{prop:tailsensbiv} hold then this is an alternative estimator for $\xi$ and $\xi_t$. Especially for estimating $\xi_t$ this is quite convenient since the estimator does not require a whole cross-section of asset prices to estimate $\xi_t$ whereas the approach in \cite{kelly2014dynamic,kelly2014tail} does. The estimated $\xi$ will be used to assess the situations under which the $\DCoES$ can be preferred to the $\DCov$. The statistical properties of the estimators and their sensitivity to outliers will be assessed via simulations in Section \ref{SSec:simresults}.\\

\noindent A simple way to use these estimators in a time-dependent manner is is to estimate them in a rolling or expanding window. While this approach does ignore the conditional mean and variance structure it models the dependence structure in a more flexible way than existing copula-GARCH by \cite{chen2006estimation,JONDEAU2006827} or DCC-GARCH by \cite{engle2002dynamic} models do. Also, in most GARCH specifications the conditional $\VaR$ and $\ES$ are linear functions of the conditional mean and variance structure the $\Delta$ measures cancel out the former while the $\xi$ estimator will also cancel out the latter. Furthermore, the result in proposition \ref{prop:robdcoes} applies to simple empirical estimators. To avoid the copula misspecification problem as much as possible, to provide a simple estimator on time-series data and to clearly show the effects and implications of propositions \ref{prop:tailsensuni} and \ref{prop:robdcoes} the simple approach is used.  More concretely, in Section \ref{Ssec:condres} the risk measures and $\xi$ are estimated daily using the rolling window of 2000 observations over the period 9th of August 2007 to 15th of September 2008. The chosen $Y$ institutions are Lehman Brothers and JP Morgan Chase to provide a clear contrast between an institution that has survived the GFC versus one that went bankrupt. For the $X$ institutions all 72 remaining institutions are chosen and all results are averaged daily over these institutions. The dates chosen correspond to the start of the subprime mortgage crisis and the bankruptcy of Lehman Brothers respectively. The setting is meant to emulate an agent at the time of the GFC receiving daily returns information and updating the model daily to assess systemic risk. The questions that arise then are: "Could these estimators have detected the impending problems with Lehman Brothers on time?" and "Do the estimators also find increasing systemic risk for an institution that survived the GFC?".  
\subsection{Simulation setup}\label{Ssec:simsetup}
To assess the empirical validity of the theoretical results in sections \ref{Ssec:tails} and \ref{Ssec:robust} and the statistical performance of all the estimators Monte-Carlo simulations will be used. To test proposition \ref{prop:tailsensuni} empirically and the statistical performance of the estimators from section \ref{Ssec:estim} the following simulation model is used. The data are drawn from Gumbel copula with $\theta=2.22\dots$ and T-distributed margins with $\mu=0,\sigma=1,\nu=3$. The usage of the Gumbel copula for joint large losses is popular in the actuarial and risk management literature and used by \cite{embrechts2001correlation,mainik2014dependence,CopulaGarchCovar}. Also, the Gumbel copula emulates the setup of proposition \ref{prop:tailsensbiv}. The value for $\theta$ is obtained from \cite{CopulaGarchCovar} where a Kendall's Tau of 0.55 is within the range of their results. From this Tau using methods described in \cite{nelsen2007introduction} the $\theta$ value is computed. The $\nu$ parameter of the T-distribution is chosen in accordance with the literature on the power-law tails of stock returns. This setup implies the following set of true values:
\begin{itemize}
    \item $\DCov_{0.95,0.95}(Y\mid X)=5.071827$
    \item  $\DCoES_{0.95,0.95}(Y\mid X)=7.383257$
    \item $\omega(0.95,0.95,C)=0.9974727$
    \item $\DCoES_{0.95,0.95}(Y\mid X)/\DCov_{0.95,0.95}(Y\mid X)=1.455739$
    \item $\xi=0.3130637$
\end{itemize}
\noindent The simulation setup consists of drawing $m=10000$ datasets of sizes $n=\{500,1000,2000,5000,10000,20000\}$ from the copula. The margins are then transformed to be T-distributed with $\nu=3$. Then, this data is transformed into a uniform distribution with the method described in \cite{hofert2019elements}. On this sample the empirical beta copula is estimated and the equation $\hat{\bar{C}}(\alpha,\omega)=(1-\alpha)(1-\beta)$ is solved using a root finding algorithm \footnote{ The default uniroot function in R is used for this with lower bound 0 upper bound 1 and tolerance $1\cdot 10^{-8}$ } to obtain $\hat{\omega}$. Applying the theoretical results the $\hat{\DCov}_{0.95,0.95}(Y\mid X)$,$\hat{\DCoES}_{0.95,0.95}(Y\mid X)$ and $\hat{\xi}$ are computed. To ensure replicability the seed within each dataset is set to fixed value. Therefore, even if the sample size of one dataset increases the data will be drawn using the same settings but across datasets the seed will differ to obtain sufficient variability. In section \ref{SSec:simresults} the results are discussed and the bias, variance and MSE are provided.  The simulation could also have been done for higher quantiles but seeing as 0.95 is already well-used in the literature, the results only become worse for higher quantiles the simulations were only performed at the 0.95 level and the GPD approximation already works well as the true $\xi$ is close to the actual $\xi$ of the distribution which is 1/3. Lastly, the 0.95 level already presents a challenge as the true $\omega$ in this scenario is already very close to its comonotonic upper bound of 0.9975. 
\\

\noindent To test proposition \ref{prop:robdcoes} empirically the following simulation model is used. The data are drawn from Gumbel copula with $\theta=2.22\dots$ and T-distributed margins with $\mu=0,\sigma=1,\nu=3$. The outlier consists of a single point $(l,l)$ in the following interval of quantiles $[0.94,0.999999]^2$ in steps of 0.000001. The lower bound is chosen as it is just below the significance of 0.95 to assess the behavior when the outlier is below the quantile level. The upper bound is very close to 1 but not equal as the support of the T-distribution is unbounded. The simulation setup then consists of drawing for each $(l,l)$ a sample of 5000 observations from the Gumbel copula. As before the margins are then transformed to be standardized T-distributed with $\nu=3$. At this stage the point $(l,l)$ is added to the dataset. Then, the observations are transformed to be uniformly distributed, the empirical copula and all relevant quantities are estimated. For each $(l,l)$ the seed is kept to he same fixed value to obtain datasets that only differ in the additional coordinate $(l,l)$. According to the results in section \ref{Ssec:robust} when taking all estimated quantities, subtracting the true value and graphing the difference versus $l$ for the $\DCov$ the estimates should be a piecewise constant function whereas for the $\DCoES$ the estimates should be piecewise linear with an intermediate area in which the estimates decrease and an extreme area where they increase in function of $l$. Based on these results it is hypothesized that the estimates of $\xi$ also have a range of $l$ where they are decreasing and for larger $l$ a range where they are increasing. Since $\xi$ is a non-linear function of the ratio of $\DCoES/\DCov$ the effects are hypothesized to be non-linear too. Although the theoretical results apply to univariate outliers in the distribution of $Y$ bivariate outliers are used to also assess the effect of bivariate outliers on estimating $\omega$. 

\subsection{Backtesting}\label{SSec:backtest}
From the representations of the $\CoV$ and $\CoES$ it can be seen that the former has two sources of variation while the latter has three. These sources of variation are: the model for the $\VaR$, the model for the dependence structure and the model for the $\ES$. The second source affects the choice of quantile at which the model should be evaluated. As this is unknown and the true $\VaR$ model are unknown it is impossible to tell if deviations are due to errors in one or the other. The issue is compounded in the case of the $\ES$ where also model error of the $\ES$ brings an additional source of variation.  Therefore, according to definitions of elicitability and identifiability as provided by \cite{gneiting2011backtest} these measures cannot be backtested. A proof of this is provided in \cite{fissler2021backtesting}. However, using the notion of joint elicitability from \cite{acerbi2014back} and multiple objective elicitability from \cite{fissler2021backtesting} it is possible to backtest risk measures with multiple sources of variation. In the literature these tests have been explored in \cite{acerbi2014back,fissler2021backtesting,banulescu2021backtesting,deng2021backtesting}. Practitioners should therefore be wary and use these newer tests in order to properly backtest the $\CoV,\CoES,\MES$ and the $\DCov/\DCoES$. Lastly, \cite{acerbi2014back} argue and prove that the $\ES$-based measures and their respective backtests are more informative to regulators and risk managers because these backtests test the amount of violations and the severity of violations whereas backtests for the $\VaR$ only consider the amount of violations. 
\section{Data}\label{Sec:data}
The dataset used for this paper is daily equity data of US financial institutions from CRSP. The dataset spans a time period from 31-12-1970 to 31-12-2020. Therefore, the dataset includes plenty of crises and rare events such as Black Monday, the Dotcom Bubble, the Great Financial Crisis and most recently the Covid crisis. Next to these aggregate shocks the data also contain some idiosyncratic shocks. In line with \cite{AdrianBrunnCovar} a financial institution is defined as a firm having an SIC-code between 6000 and 6800. In line with the literature on asset pricing ADRs, SDIs and REITs are excluded. Furthermore, missing returns and prices (as well as prices lesser than or equal to zero) and firms with less than 260 weeks of returns data are excluded.  Only active firms are kept in the sample. The CRSP data is also merger adjusted in a sense that at any given time $t$ only firms that were not acquired until $t$ are in the dataset. Through the PERMNO identifier firms that have changed name, SIC code or even stock ticker over time are tracked. It must be noted that its fairly rare for a firm to change SIC codes, a few prominent examples include Goldman Sachs changing to the 6730 (bank holding firm) SIC code during the GFC and VISA and Mastercard changing to the 7389 SIC code (Business services, not elsewhere classified). This last change has as a consequence that these two firms are kept out of the sample as there are fewer than 260 weeks of returns data between the IPO date and the SIC code change. For a full and precise list of all variables and the filtering procedures we refer the reader to Section \ref{Ssec:datafilt}. In the end a dataset of in total 6.182.652 observations over 18.613 days and over 1564 firms is obtained. Mind that since not all firms have data available for the entire time span the panel dataset is unbalanced. To analyse the data R \cite{Rsoftware} is used. For the details regarding the computer and R setup including packages, see section \ref{App:compR}. Lastly, to aid in weekly aggregation of our data a trading week is defined to consist of 5 days and a trading year to consist of 51 weeks. This division results in a total of 2523 weeks. For the weekly returns aggregation daily returns $r_i$ were aggregated according to the following formula $\prod_{i=1}^{5}(1+r_i)-1$ for day $i=1,\dots,5$ in order to avoid the instabilities faced when using daily prices as for some firms the price and outstanding share data exhibited jumps that could not be seen when looking at share data of the same firms on for example Yahoo Finance. However, returns did not seem to be subject to these anomalies and also the anomalies do not affect market value calculations as even with the jumps the total market value would not change much. \newline

\noindent The losses of the financial system index are defined as follows: \begin{mydef}\label{def:finsysloss}
Let $X^1_i,\dots,X^T_i$ be the equity returns of institution $i$ with $i=1,\dots,N$ from time $t=1,\dots,T$. Let $\text{MV}^t_i$ be the market-value of institution $i$ at time $t$. Then the loss of the financial system at time $t$ is  defined as:
\begin{equation*}
    X^t_{sys}=\frac{-\displaystyle \sum_{i=1}^{N}\text{MV}^t_i X^t_i}{\displaystyle \sum_{i=1}^{N}\text{MV}^t_i}
\end{equation*}
\end{mydef}
\noindent This definition is equivalent to the one in \cite{AdrianBrunnCovar}. The definition applies to both daily and weekly equity data.\\

\noindent The choice of financial institutions for the systemic risk analyses is based on \cite{GIRARDI20133169} but with additional institutions representing stock exchanges. These additions are:  CME Group Inc (CME), Intercontinental Exchange Inc (ICE)\footnote{ Since ICE has acquired NYSE/Euronext in 2012 the NYSE is also included. } and NASDAQ Inc (NDAQ). This results in a sample of 73 financial institutions for which the risk measures will be computed \footnote{For some institutions studied in \cite{GIRARDI20133169} the Brunnermeier dataset did not have data so these drop out. The institutions are: Leucadia International, Union Pacific, Berkshire Hathaway Inc (A and B class).}. 

\section{Results}\label{Sec:results}
\subsection{Simulation}\label{SSec:simresults}
First, the results of the extreme value simulations are provided in Table \ref{tab:Bias} and \ref{tab:variance}. These tables contain the bias and variance respectively of the estimates of $\DCov,\DCoES,\omega$ and $\xi$.
\begin{table}[H]
\parbox{.45\linewidth}{
\centering
\begin{tabular}{ccccc}
  \hline
 $n$& $\DCov$ & $\DCoES$ & $\omega$ & $\xi$ \\
  \hline
500 & -0.559 & -2.099 & $-1\cdot 10^{-4}$  & -0.305 \\ 
  1000 & -0.276 & -0.970 & $-4.2\cdot 10^{-5}$ & -0.155 \\ 
  2000 & -0.171 & -0.417 & $-2.1\cdot 10^{-5}$ & -0.077 \\ 
  5000 & -0.057 & -0.181 & $-8\cdot 10^{-6}$  & -0.038 \\ 
  10000 & -0.028 & -0.075 & $-4\cdot 10^{-6}$ & -0.019 \\ 
  20000 & -0.011 & -0.045 & $-2\cdot 10^{-6}$ & -0.011 \\ 
   \hline
\end{tabular}
\caption{Bias of the estimates}
\label{tab:Bias}
}
\hfill
\parbox{.45\linewidth}{
\centering
\begin{tabular}{ccccc}
  \hline
 $n$&$\DCov$ & $\DCoES$ & $\omega$  & $\xi$ \\ 
  \hline
  500&3.883 & 15.348 & $1.36\cdot 10^{-7}$  & 0.088 \\ 
  1000&2.169 & 14.017 & $3.9\cdot 10^{-8}$  & 0.058 \\ 
  2000&1.211 & 8.850 & $1.5\cdot 10^{-8}$  & 0.039 \\ 
  5000&0.509 & 3.903 & $5\cdot 10^{-9}$  & 0.018 \\ 
  10000&0.257 & 1.943 & $2\cdot 10^{-9}$  & 0.010 \\ 
  20000&0.131 & 0.989 & $1\cdot 10^{-9}$ & 0.006 \\ 
   \hline
\end{tabular}
\caption{Variance of the estimates}
\label{tab:variance}
}
\end{table}
\noindent In Table \ref{tab:Bias} it is clear that in finite samples all estimates are biased downwards. This bias does decrease as $n$ increases albeit at a slow rate for the risk measures because there need to be sufficient tail observations in the data which by definition are very rare and the method is fully empirical . The $\DCov$ estimates are more accurate than the $\DCoES$ estimates because fewer tail observations are required for the level of the quantile than for the mean of the distribution above said level. A consequence of this is that the variance of $\DCoES$ estimates can be quite severe.  The bias of the $\omega$ estimates quickly decreases to about 2 orders of magnitude above the tolerance of the numerical solver. This quick decrease is necessary as at these high quantiles even a small estimation error in $\omega$ could result in a large estimation error in $\DCoES$. Hence, alternatively, one could use the fully empirical method to estimate $\omega$ with high precision and then use a well-specified model for the margins to compute the risk measures with parametric precision. These results show that for reliable estimates of the risk measures one needs at least 2000 observations. The downwards bias is also a warning to practitioners that any estimate is likely to be too low. In general, these results show that proposition \ref{prop:tailsensuni} holds relevance for empirical use and that one should consider extreme value properties and methods when estimating these risk measures. A table with the MSE for each estimator can be found in the appendix. \\

\noindent Regarding the results of the outlier simulation these will be displayed graphically for a risk measure as a function of $l$. Mind that the actual levels of $l$ and not the corresponding quantiles will be used. However, as a reference the losses corresponding to the quantiles $\{0.95,0.975,0.99,0.9975\}$ are $\{2.353363,3.182446,4.540703\\,7.453319\}$. \\

\begin{figure}[H]
     \centering
     \subfigure[]{
     \includegraphics[width=0.45\textwidth]{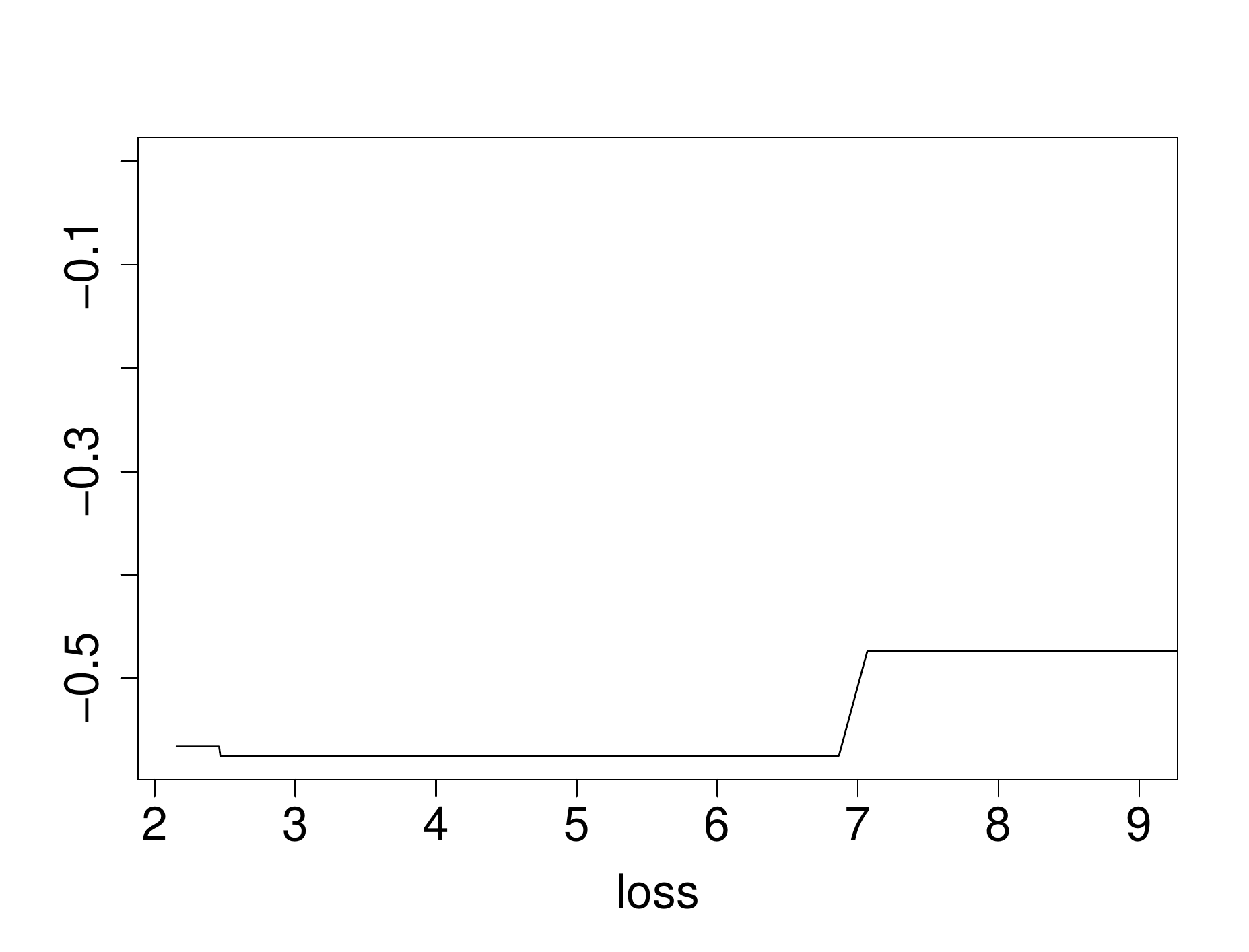}
     \label{fig:dcovrob}
     }
     \qquad
        \subfigure[]{
     \includegraphics[width=0.45\textwidth]{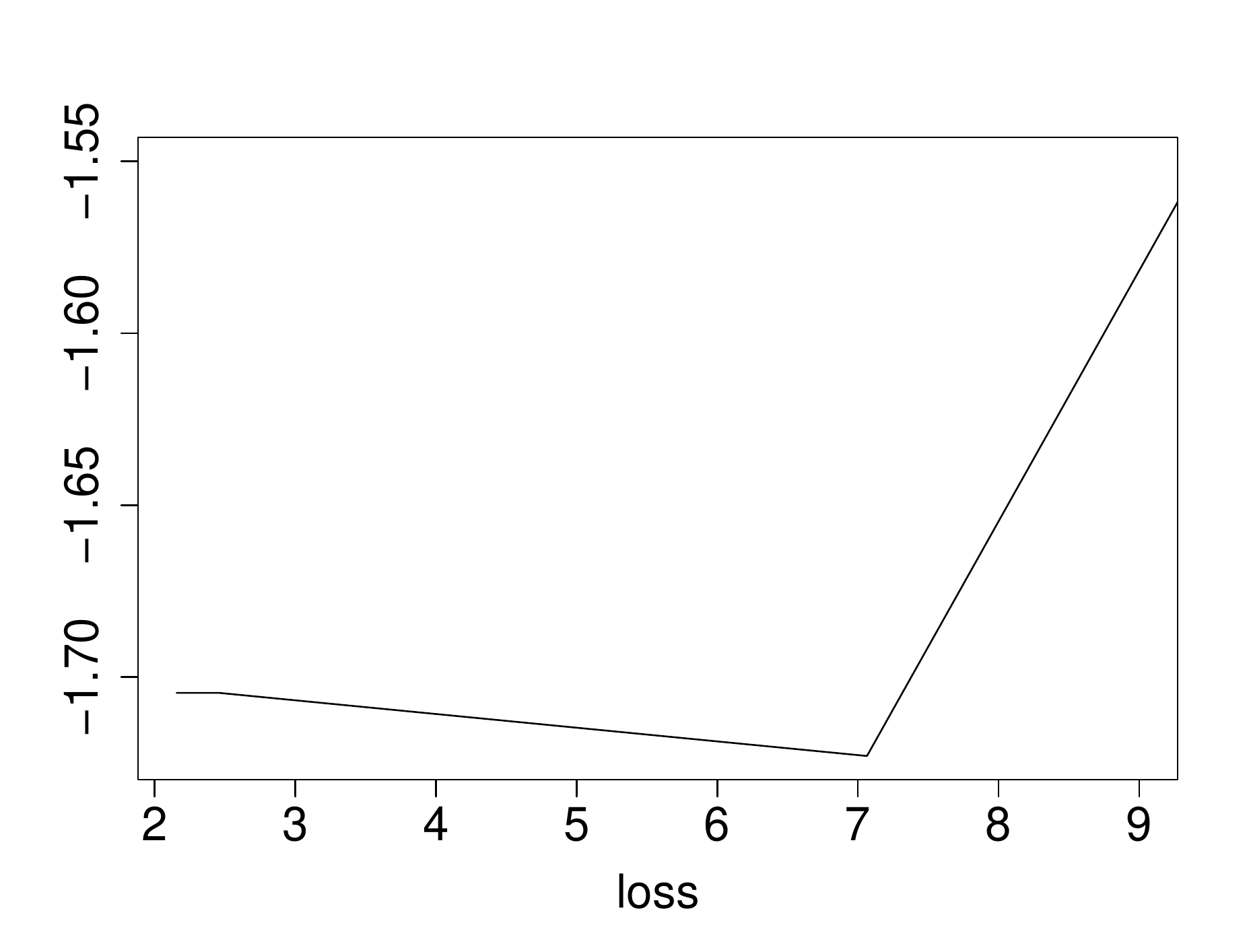}
     \label{fig:dcoesrob}
     }
     \qquad
     \subfigure[]{
     \includegraphics[width=0.45\textwidth]{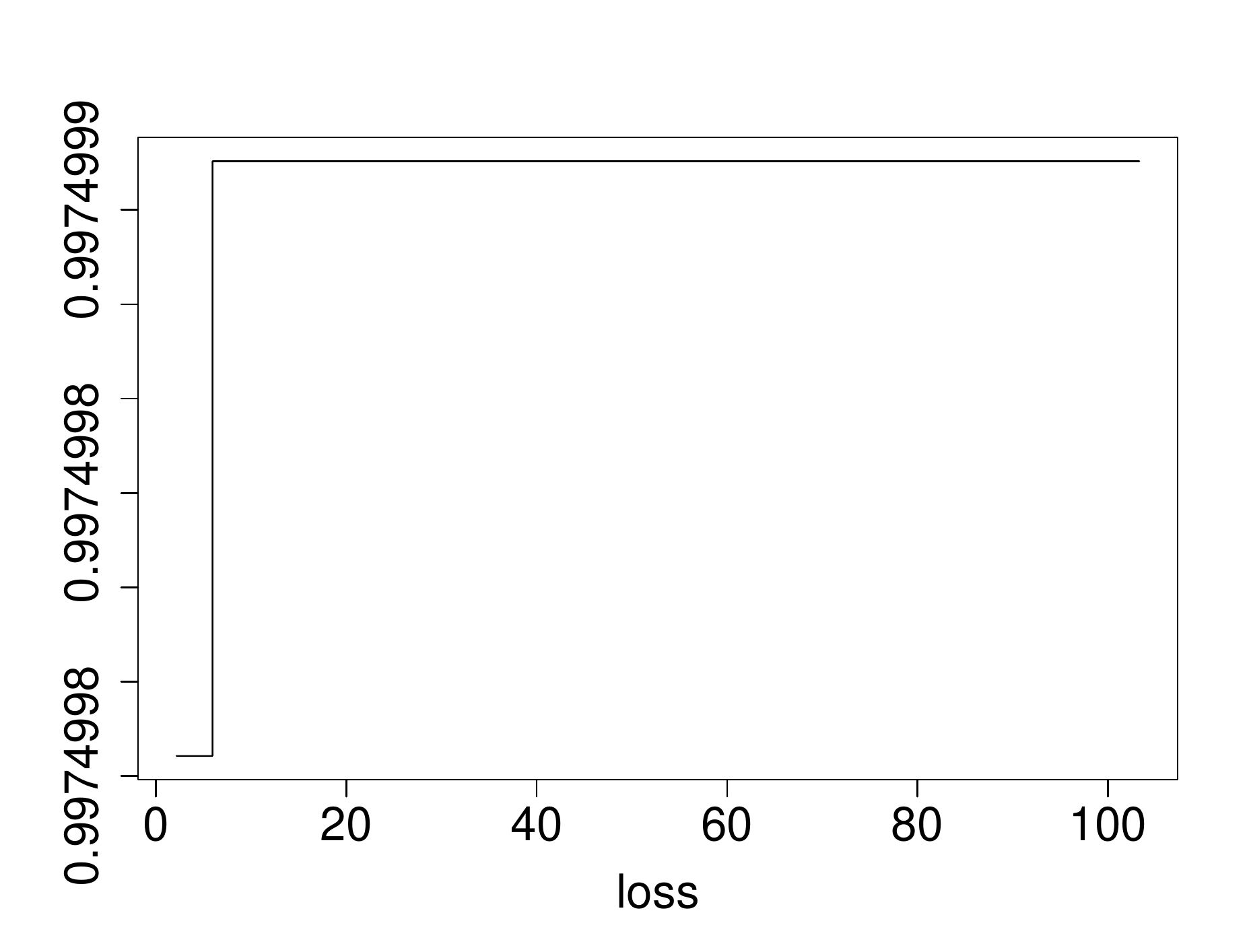}
     \label{fig:omrob}

     }
     \qquad
     \subfigure[]{
     \includegraphics[width=0.45\textwidth]{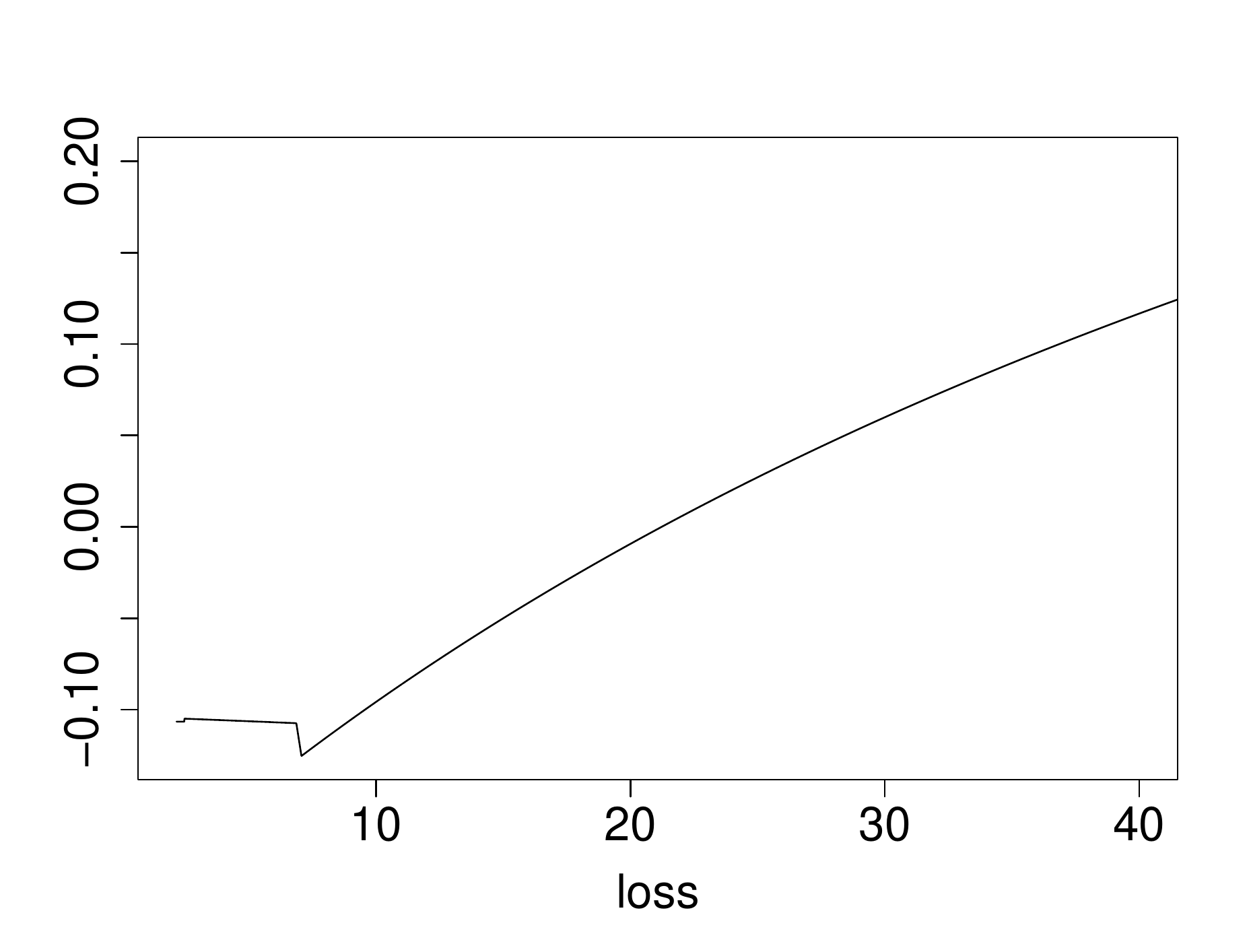}
     \label{fig:xirob}
     }
     
     \caption{Plots of the simulation on the sensitivity of risk measure estimates (from top left to bottom right: $\DCov,\DCoES,\omega$ and $\xi$) to an outlier at joint loss $(l,l)$ with $l$ on the x-axis. For all figures except \ref{fig:omrob} the y-axis shows estimates minus the true value. In figure \ref{fig:omrob} the y-axis shows the estimates.}
     \label{fig:robs}
 \end{figure}
 
 \noindent Figure \ref{fig:robs} clearly shows the estimates tend to be biased downwards with the $\DCov$ having lower bias than the $\DCoES$. However, now it becomes clear that outliers can change this. In the case of the $\DCov$ the sensitivity function is indeed as hypothesized a piecewise constant function with an intermediate region where the downward bias is larger and a region beyond it where the downward bias is smaller than even before the intermediate region. The $\DCoES$ sensitivity function confirms proposition \ref{prop:robdcoes} that it is piecewise linear with an intermediate region where the bias is decreasing in $l$ and a region beyond it where it is increasing in $l$. For sufficiently large $l$ this upward effect would make the the estimate unbiased while the $\DCov$ estimates stay constant after the jump upwards from the intermediate region. Due to the use of a joint outlier $(l,l)$ the $\omega$ estimates are also affected but only to a very small extent. Interestingly, the sensitivity function looks very much like that of the $\VaR$. The sensitivity function of the $\xi$ estimates seems to confirm the hypothesis made in section \ref{Ssec:simsetup} and has an intermediate region where it sharply decreases before non-linearly increasing in $l$. These results provide an additional warning for practitioners: sharply decreasing estimates of these risk measures might be due to intermediate losses building up which signals a build up of risk rather than a reduction! For very large losses $\DCoES$ and $\xi$ estimates are sensitive to the size of the loss. Hence, for these estimators sharply decreasing values could be followed up by sharply increasing values as the built up risk manifests as losses beyond the intermediate range. Lastly, in a great confirmation of the theoretical results the points as which the different pieces of the $\DCoES$ begin and end are, up to sampling error of $\omega$, exactly where proposition \ref{prop:robdcoes} indicated they would be.   
\subsection{Data analysis: unconditional case}\label{Ssec:uncondres}
In this section, the measures will be estimated in an unconditional fashion. First, the unconditional power-law properties of the financial system and financial institution returns will be checked to see if they conform with the broader empirical literature and assess the change if the return interval is changed from daily to weekly. \\

\noindent First, by examining the unconditional tails of the loss distribution of the financial system index, the entire dataset and of the sample of chosen institutions at the daily and weekly interval the power-law results by \cite{PL5,PL6} are confirmed with estimates of $\xi$ in $[0.3296567,0.3971911
]$ so therefore well around the value of $1/3$. In Table \ref{tab:sumstats1day} and \ref{tab:sumstats1week} the summary statistics for the risk measures computed on the sample of 73 institutions versus the system index on daily and weekly returns data are provided. 
\begin{table}[H]
    \centering
    \begin{tabular}{|c|c|c|c|c|c|c|c|}\hline
         &Mean&Median&SD&Q1&Q3&Min&Max  \\ \hline
         $\DCov_{0.95,0.95}$&0.04865&0.05318&0.0110627&0.04349&0.05527&0.01720&0.06947  \\\hline
         $\DCoES_{0.95,0.95}$&0.05937&0.06328&0.0110131&0.05972&0.06483&0.01780&0.07052  \\\hline
         $\omega(0.95,0.95,C)$&0.9971&0.9973&0.0007990754&0.9971&0.9974&0.9920&0.9975\\\hline
         Ratio& 1.236& 1.196&0.1301443& 1.150& 1.364&1.015&1.644 \\\hline
         $\xi$&0.18258&0.16373&0.07967325&0.13027&0.26676&0.01482&0.39158 \\\hline
    \end{tabular}
    \caption{Summary statistics of daily $\DCov$,$\DCoES$, $\omega$, the ratio $\frac{\DCoES}{\DCov}$ and the $\xi$ implied by the ratio at $\alpha=\beta=0.95$. For $\alpha=\beta=0.95$ one gets that $\omega_{\text{min}}=0.95$ and $\omega_{\text{max}}=0.9975$.}
    \label{tab:sumstats1day}
\end{table}
\begin{table}[H]
    \centering
    \begin{tabular}{|c|c|c|c|c|c|c|c|}\hline
         &Mean&Median&SD&Q1&Q3&Min&Max  \\ \hline
         $\DCov_{0.95,0.95}$&0.10118&0.10957 & 0.0312604&0.08858&0.12225&0.03254&0.14469  \\\hline
         $\DCoES_{0.95,0.95}$&0.13656&0.14992&0.03152259&0.13314&0.15226&0.02801&0.17574  \\\hline
         $\omega(0.95,0.95,C)$&0.9971& 0.9975&0.0008783222&0.9972&0.9975&0.9935&0.9975\\\hline
         Ratio& 1.4099& 1.3465&0.3526316&1.1958&1.4735&0.8408& 2.8606 \\\hline
         $\xi$&0.2593&0.2573&0.1371591&0.1638&0.3213&-0.1894&0.6504 \\\hline
    \end{tabular}
    \caption{Summary statistics of weekly $\DCov$, $\DCoES$,  $\omega$, the ratio $\frac{\DCoES}{\DCov}$ and the $\xi$ implied by the ratio at $\alpha=\beta=0.95$. For $\alpha=\beta=0.95$ one gets that $\omega_{\text{min}}=0.95$ and $\omega_{\text{max}}=0.9975$.}
    \label{tab:sumstats1week}
\end{table}
\noindent The results show that on both for daily and weekly losses the $\omega$s are very close to the upper comonotonic bound ($\omega(0.95,0.95,C)_{\text{max}}=0.9975$) which shows that the risk measures are computed at very high quantiles. This result should make practitioners use very large samples and appropriate estimators in order to makes sure results are reliable. For higher $\alpha=\beta$ this becomes even worse. Therefore, in the rest of this paper all measures are computed at $\alpha=\beta=0.95$. Additionally, all measures are computed at the daily level because even at the lowest quantile the median amount of observations used for $\CoES$ estimation on weekly data was 5 while on the daily data this was 24. Therefore, in order to maximize the stability of estimates the sample of daily data is used. As predicted in section \ref{Ssec:aggr} due to using a weighted average of losses the $\DCov$ and $\DCoES$ do not show much heterogeneity between institutions and also the differences between the risk measures are small. There is a bit more variety in the ratios and the $\xi$ estimates. This is not surprising as also with the traditional methods the estimated tail coefficient can be highly dependent on the choice of threshold. In order to better test the results and implications of section \ref{Ssec:aggr} next the network version of these measures are computed across the full grid of $73\cdot72=5256$ ordered pairs of different institutions.\newpage \noindent In Table \ref{tab:sumstats1daynetw} the summary statistics for the network estimation are provided.  
\begin{table}[H]
    \centering
    \begin{tabular}{|c|c|c|c|c|c|c|c|}\hline
         &Mean&Median&SD&Q1&Q3&Min&Max  \\ \hline
         $\DCov_{0.95,0.95}$&0.05911531&0.05675903&0.02533318&0.04193931&0.07294496&0.004581243&0.2825005  \\\hline
         $\DCoES_{0.95,0.95}$&0.07916396&0.07277265&0.03811298&0.05443875&0.1012622&0.006456835&0.5619791  \\\hline
         $\omega(0.95,0.95,C)$&0.9948876&0.9960929&0.003585537&0.9946132&0.9967199&0.9631791&0.9975\\\hline
         Ratio&1.345733&1.289713&0.2848762&1.183651&1.435364&0.6777686&5.599564 \\\hline
         $\xi$&0.2306807&0.2246334&0.1338684&0.1551567 &0.3033125&-0.4754298&0.8214147 \\\hline
    \end{tabular}
    \caption{Summary statistics of daily network $\DCov$,$\DCoES$, $\omega$, the ratio $\frac{\DCoES}{\DCov}$ and the $\xi$ implied by the ratio at $\alpha=\beta=0.95$. For $\alpha=\beta=0.95$ one gets that $\omega_{\text{min}}=0.95$ and $\omega_{\text{max}}=0.9975$.}
    \label{tab:sumstats1daynetw}
\end{table}
\noindent The table shows that the results from section \ref{Ssec:aggr} are confirmed because not just do both risk measures show more variability compared to the systemic version but also the standard deviation of the $\DCoES$ estimates is almost twice that of the $\DCov$ estimates again showing the higher sensitivity. Similar conclusions hold for the ratios and the $\xi$ estimates, There are some very low values of $\DCov,\DCoES$ and $\omega$ but these can mostly be attributed to the institution Commerce Bancorp Inc NJ (CBH) which was acquired on October 2nd 2007 by the Toronto Dominion bank. Hence, the extremes of the GFC are not present in the CBH stock loss data. For this reason, the ratio and $\xi$ summary statistics are computed with CBH omitted. Notice that some $\xi$ estimates are negative. This could be due to the true $\xi$ being close to zero combined with the downward bias of the estimator and some intermediate losses that push the estimates further down. The $\omega$ (see Figure \ref{fig:ommat}, Appendix) estimates give credence to the notion that in times of a financial market crisis (almost) everything moves in the same direction. Also, the $\omega$ estimates for the stock exchanges and broker-dealers are higher than for other intermediaries. Besides these patterns the $\omega$ estimates are quite uniform. Therefore, most of the variation in the risk measures can be attributed to the tails of the losses of $Y$. 
\begin{figure}[H]
    \centering
    \includegraphics[width=1.2\textwidth]{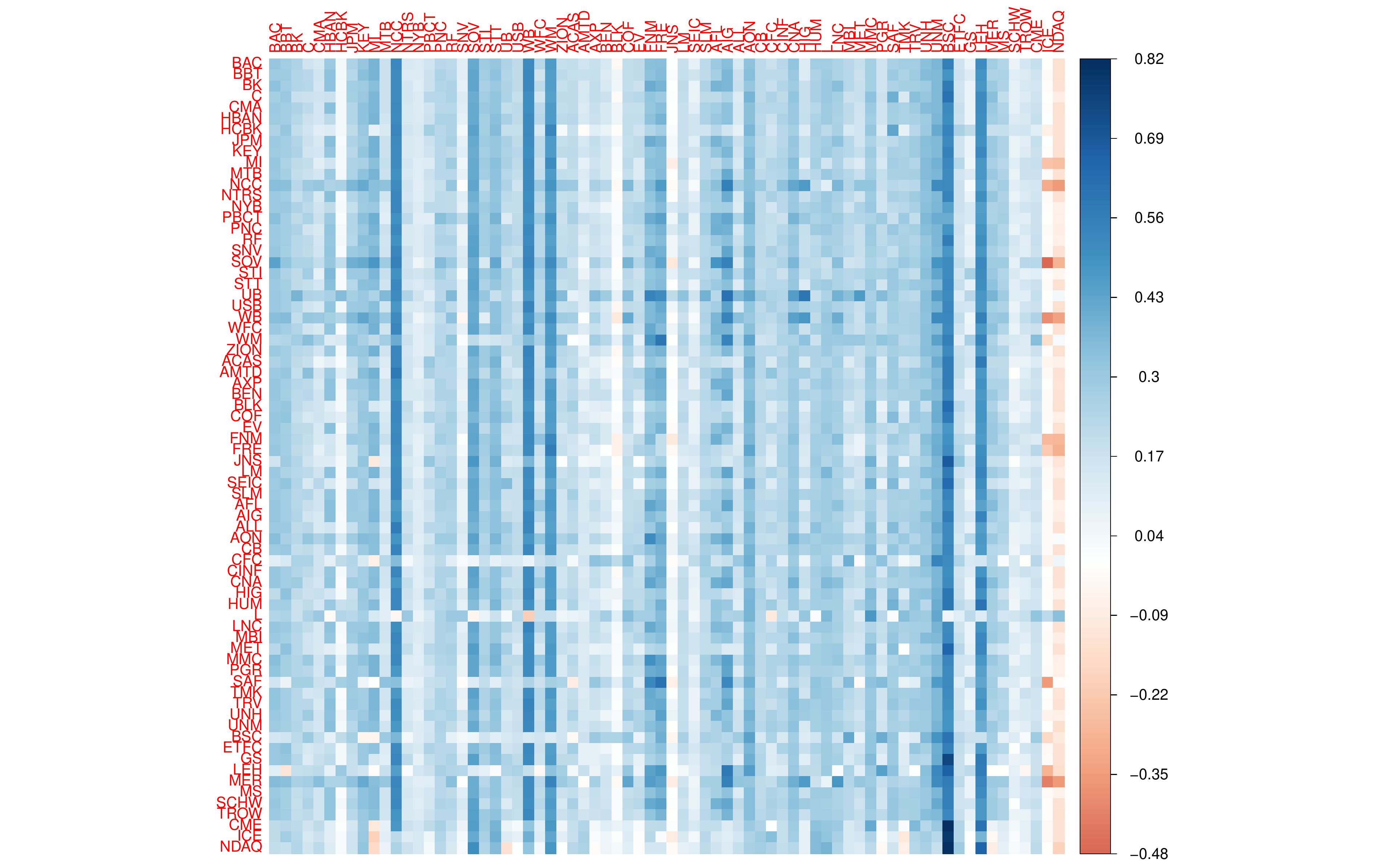}
    \caption{Daily network $\xi$ estimates for the network of institutions. $X$ institutions are on the rows, $Y$ institutions are on the columns. CBH has been omitted for reasons mentioned above. }
    \label{fig:netwxi}
\end{figure}

\noindent In Figure \ref{fig:netwxi} it can be seen that the estimates for each $Y$ institution show a clear band structure with estimates of the same $Y$ having similar magnitudes. Because the $\omega$ estimates across all pairs are rather similar this difference is mostly due to the difference in the tails of $Y$ as shown in proposition \ref{prop:tailsensuni}. The differences within the estimates of a $Y$ institution can be explained by joint outliers that can push up $\omega$ estimates. This confirms the effects of propositions \ref{prop:tailsensbiv} and \ref{prop:robdcoes}. In these plots it is rather straightforward to identify the institutions that failed or were close to failing during the GFC as these tend to exhibit elevated estimated when being the $Y$ institution. Notable examples include: National City Corp (NCC), Wachovia Bank (WB), Washington Mutual (WM), Fannie Mae (FNM), Freddie Mac (FRE) ,AIG,Bear Staerns (BSC) and Lehman Brothers (LEH). Finally, it seems that the stock exchanges exhibit some particularly strong links with all the examples mentioned with extreme estimates if Bear Stearns and Lehman Brothers were the $Y$ institution. Even more striking is that the link in the opposite direction is rather weak with the $\xi$'s implying they have (sub)-Gaussian tails. Some caution must be taken here as the negative $\xi$ estimates are likely a combination of the true $\xi$ being close to zero, the negative bias of the $\xi$ estimator and some intermediate losses pushing the estimates further downwards. So even if the tails might be well-behaved there could still be some sizeable losses. These findings show that one cannot ignore the role of stock exchanges in systemic risk analyses and provide evidence that exchanges as a financial intermediary have a sizable impact on other intermediaries if the exchanges are under stress. Hence, the $\xi$ estimates contain information that the $\omega$ estimates do not have: information on exposure and the effect of the dependence. 
\subsection{Data analysis: conditional case}\label{Ssec:condres}
In this section the risk measures and $\xi$ are estimated daily using the rolling window of 2000 observations over the period 9th of August 2007 to 15th of September 2008. The chosen $Y$ institutions are Lehman Brothers and JP Morgan Chase to provide a clear contrast between an institution that has survived the GFC versus one that went bankrupt. For the $X$ institutions all 72 remaining institutions are chosen and all results are averaged daily over these institutions. All estimates are at the 95\% level.
\begin{figure}[H]
     \centering
     \subfigure[]{
     \includegraphics[width=0.65\textwidth]{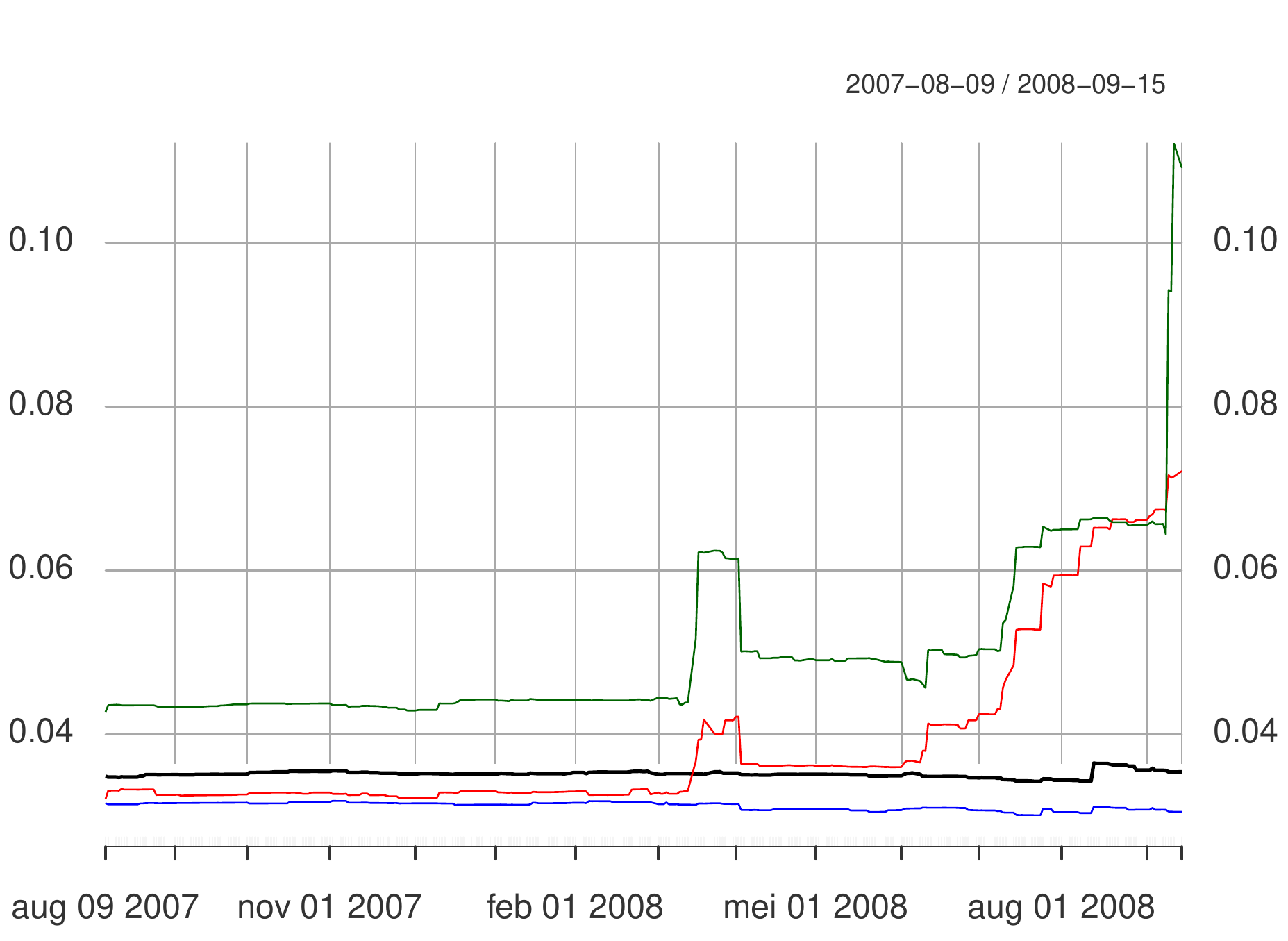}
     \label{fig:bankrisk}
     }
     \qquad
        \subfigure[]{
     \includegraphics[width=0.65\textwidth]{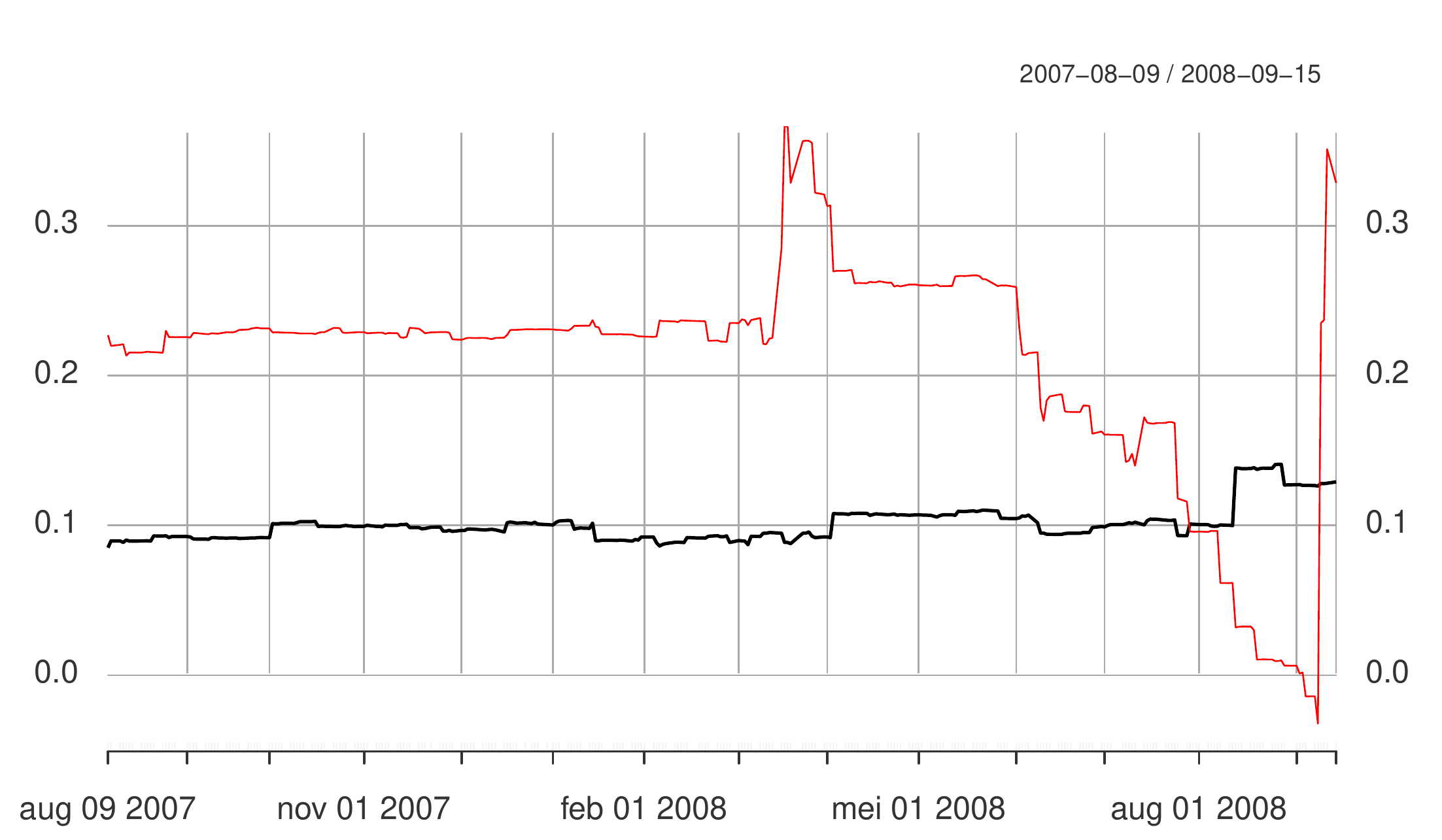}
     \label{fig:xiplot}
     }

     \caption{Figure \ref{fig:bankrisk} contains the $\DCov$ and $\DCoES$ estimates for JP Morgan Chase ($\DCov$:Blue, $\DCoES$:Black) and Lehman Brothers ($\DCov$:Red, $\DCoES$:Green) averaged over all the other institutions. Hence, the estimates show the average risk contributions of other institutions to JP Morgan Chase and Lehman Brothers. Figure \ref{fig:xiplot} shows the $\xi$ estimates for JP Morgan Chase (Black) and Lehman Brothers (Red).  }
     \label{fig:condmeas}
 \end{figure}\newpage
 \noindent Figure \ref{fig:condmeas} shows that over the whole period the $\DCoES$ and $\DCov$ estimates consider Lehman Brothers to be more sensitive to shocks at other institutions than JP Morgan Chase. The $\DCoES$ and $\DCov$ for both show similar patterns but as expected due to the sensitivity function of the $\DCoES$ the estimates of the $\DCoES$ of Lehman Brothers quickly jump upwards some months before bankruptcy but also in the few days before bankruptcy jump to before unseen levels. Hence, the notion that the $\DCoES$ could function as an early warning indicator seems to hold. In contrast, the estimates of both risk measures of JP Morgan Chase barely change apart from around a month before the Lehman Brothers bankruptcy. This can be interpreted as another sign that risk is building up in other parts of the financial system instead of just in Lehman Brothers. The negative reaction to intermediate losses of the $\DCoES$ and $\DCov$ can also be seen in the estimates of Lehman Brothers right before the large spike at the end. Again, this behavior could be used as an early-warning sign. The $\xi$ estimates show an even more stark difference with those of JP Morgan Chase slowly trending up and a jump around mid-August. In stark contrast, the estimates of Lehman Brothers spike mid-March, rapidly decrease from June to the start of September and spike again just a few days (5) before bankruptcy. This pattern is rather interesting as it is due to the $\DCoES$ and $\DCov$ estimates becoming more similar. This could be due to the $\DCov$ catching up to the $\DCoES$ as large losses are accumulated which temporarily creates the very false impression that the tail of Lehman Brothers is becoming more Gaussian and hence less risky. However, looking at the actual levels of both risk measures its clear that even if the tail seems to become lighter the distribution has shifted upwards (in a first- or second-order dominance sense) as well which still reflects the increased risk. Because $\DCov$ and $\DCoES$ estimates are necessary for the $\xi$ estimator of this paper practitioners are advised to always look at both, compare and to be wary of strongly fluctuating $\xi$ estimates. Therefore, highly volatile $\xi$ estimates can serve as an early-warning sign. Lastly, the hypothesized time-series properties of the $\DCov$ and $\DCoES$ seem to hold as the estimates of the former are more similar over time while the estimates of the latter are more volatile. Additional evidence for this claim is provided by ACF plots in figure \ref{fig:acfs} in the appendix.   
\section{Conclusion}\label{Sec:conc}
Based on a univariate representation several statistical properties of the $\DCov,\DCoES$ and $\MES$ are explored. This leads to novel empirical estimators for the $\DCov,\DCoES$ and the power-law coefficient. The theoretical exploration also highlights the importance of extreme value theory, outliers and their effects on the statistical behavior of these risk measures. Empirically, these theoretical results are confirmed, the statistical performance of the novel estimators is assessed and the novel methods are applied to an extended version of the dataset of \cite{AdrianBrunnCovar}. The findings are that the $\MES$ is not suitable for measuring (joint) extreme risk, under aggregation of $Y$ the difference between the $\DCov$ and $\DCoES$ is more marginal, the $\DCoES$ is most suited for network risk, the extremes matter when estimating the risk measures, outliers can greatly affect estimates which can work as an early-warning system, for accurate estimation very large samples sizes are necessary, the risk measures generally are underestimated and the power-law coefficient estimator shows its merits when applied to a case-study on financial data.  
\bibliographystyle{apalike}
\bibliography{ref}

\begin{thebibliography}{}

\bibitem[Acerbi et~al., 2001]{acerbiES1}
Acerbi, C., Nordio, C., and Sirtori, C. (2001).
\newblock Expected shortfall as a tool for financial risk management.
\newblock {\em arXiv preprint cond-mat/0102304}.

\bibitem[Acerbi and Szekely, 2014]{acerbi2014back}
Acerbi, C. and Szekely, B. (2014).
\newblock Back-testing expected shortfall.
\newblock {\em Risk}, 27(11):76--81.

\bibitem[Acerbi and Tasche, 2002]{ACERBIES2}
Acerbi, C. and Tasche, D. (2002).
\newblock On the coherence of expected shortfall.
\newblock {\em Journal of Banking \& Finance}, 26(7):1487--1503.

\bibitem[Acharya et~al., 2016]{MES1}
Acharya, V.~V., Pedersen, L.~H., Philippon, T., and Richardson, M. (2016).
\newblock {Measuring Systemic Risk}.
\newblock {\em The Review of Financial Studies}, 30(1):2--47.

\bibitem[Adrian and Brunnermeier, 2016]{AdrianBrunnCovar}
Adrian, T. and Brunnermeier, M.~K. (2016).
\newblock {CoVaR}.
\newblock {\em American Economic Review}, 106(7):1705--41.

\bibitem[Artzner et~al., 1999]{Cohriskmeas}
Artzner, P., Delbaen, F., Eber, J.-M., and Heath, D. (1999).
\newblock Coherent measures of risk.
\newblock {\em Mathematical Finance}, 9(3):203--228.

\bibitem[Balkema and de~Haan, 1974]{balkdehaan}
Balkema, A.~A. and de~Haan, L. (1974).
\newblock Residual life time at great age.
\newblock {\em The Annals of Probability}, 2(5):792--804.

\bibitem[Banulescu-Radu et~al., 2021]{banulescu2021backtesting}
Banulescu-Radu, D., Hurlin, C., Leymarie, J., and Scaillet, O. (2021).
\newblock Backtesting marginal expected shortfall and related systemic risk
  measures.
\newblock {\em Management Science}, 67(9):5730--5754.

\bibitem[Barro, 2006]{barro2006rare}
Barro, R.~J. (2006).
\newblock Rare disasters and asset markets in the twentieth century.
\newblock {\em The Quarterly Journal of Economics}, 121(3):823--866.

\bibitem[Beck et~al., 2020]{beck2020bank}
Beck, T., Radev, D., and Schnabel, I. (2020).
\newblock Bank resolution regimes and systemic risk.

\bibitem[Benoit et~al., 2017]{benoit2017risks}
Benoit, S., Colliard, J.-E., Hurlin, C., and P{\'e}rignon, C. (2017).
\newblock Where the risks lie: A survey on systemic risk.
\newblock {\em Review of Finance}, 21(1):109--152.

\bibitem[Bernard and Czado, 2015]{BERNARD2015104}
Bernard, C. and Czado, C. (2015).
\newblock {Conditional quantiles and tail dependence}.
\newblock {\em Journal of Multivariate Analysis}, 138:104--126.
\newblock High-Dimensional Dependence and Copulas.

\bibitem[Bernardi et~al., 2017]{BERNARDI20178}
Bernardi, M., Durante, F., and Jaworski, P. (2017).
\newblock {CoVaR of families of copulas}.
\newblock {\em Statistics \& Probability Letters}, 120:8--17.

\bibitem[BIS, 2014]{FTRB}
BIS (2014).
\newblock {Fundamental review of the trading book: A revised market risk
  framework}.
\newblock \url{https://www.bis.org/publ/bcbs265.pdf}.

\bibitem[Bisias et~al., 2012]{bisias2012survey}
Bisias, D., Flood, M., Lo, A.~W., and Valavanis, S. (2012).
\newblock A survey of systemic risk analytics.
\newblock {\em Annu. Rev. Financ. Econ.}, 4(1):255--296.

\bibitem[Brownlees and Engle, 2016]{MES2}
Brownlees, C. and Engle, R.~F. (2016).
\newblock {SRISK: A Conditional Capital Shortfall Measure of Systemic Risk}.
\newblock {\em The Review of Financial Studies}, 30(1):48--79.

\bibitem[Brunnermeier et~al., 2020]{brunnermeier2020asset}
Brunnermeier, M., Rother, S., and Schnabel, I. (2020).
\newblock Asset price bubbles and systemic risk.
\newblock {\em The Review of Financial Studies}, 33(9):4272--4317.

\bibitem[Chen, 2007]{NonparESest}
Chen, S.~X. (2007).
\newblock {Nonparametric Estimation of Expected Shortfall}.
\newblock {\em Journal of Financial Econometrics}, 6(1):87--107.

\bibitem[Chen and Fan, 2006]{chen2006estimation}
Chen, X. and Fan, Y. (2006).
\newblock {Estimation and model selection of semiparametric copula-based
  multivariate dynamic models under copula misspecification}.
\newblock {\em Journal of econometrics}, 135(1-2):125--154.

\bibitem[Cirillo and Taleb, 2016]{GPDtrick}
Cirillo, P. and Taleb, N.~N. (2016).
\newblock Expected shortfall estimation for apparently infinite-mean models of
  operational risk.
\newblock {\em Quantitative Finance}, 16(10):1485--1494.

\bibitem[Claessens, 2015]{Macpruclaessens}
Claessens, S. (2015).
\newblock An overview of macroprudential policy tools.
\newblock {\em Annual Review of Financial Economics}, 7(1):397--422.

\bibitem[Cont, 2001]{ContFinanceempirics}
Cont, R. (2001).
\newblock Empirical properties of asset returns: stylized facts and statistical
  issues.
\newblock {\em Quantitative Finance}, 1(2):223--236.

\bibitem[Cont et~al., 2010]{ContRobrisk}
Cont, R., Deguest, R., and Scandolo, G. (2010).
\newblock Robustness and sensitivity analysis of risk measurement procedures.
\newblock {\em Quantitative Finance}, 10(6):593--606.

\bibitem[Dahl et~al., 2019]{Rxtable}
Dahl, D.~B., Scott, D., Roosen, C., Magnusson, A., and Swinton, J. (2019).
\newblock {\em xtable: Export Tables to LaTeX or HTML}.
\newblock R package version 1.8-4.

\bibitem[Danielsson et~al., 2005]{danielsson2005subadditivity}
Danielsson, J., Jorgensen, B.~N., Mandira, S., Samorodnitsky, G., and De~Vries,
  C.~G. (2005).
\newblock {Subadditivity re-examined: the case for Value-at-Risk}.
\newblock Technical report, Cornell University Operations Research and
  Industrial Engineering.

\bibitem[Dan{\'\i}elsson et~al., 2013]{danielsson2013fat}
Dan{\'\i}elsson, J., Jorgensen, B.~N., Samorodnitsky, G., Sarma, M., and
  de~Vries, C.~G. (2013).
\newblock {Fat tails, VaR and subadditivity}.
\newblock {\em Journal of econometrics}, 172(2):283--291.

\bibitem[Danielsson et~al., 2001]{Danielsson2001}
Danielsson, J., Keating, C., Shin, H.~S., and Goodhart, C. (2001).
\newblock {An Academic Response to Basel II}.
\newblock Fmg special papers, Financial Markets Group.

\bibitem[Daníelsson, 2008]{DANIELSSON2008321}
Daníelsson, J. (2008).
\newblock Blame the models.
\newblock {\em Journal of Financial Stability}, 4(4):321--328.
\newblock Regulation and the Financial Crisis of 2007-08: Review and Analysis.

\bibitem[De~Haan and Resnick, 1977]{EVcop1}
De~Haan, L. and Resnick, S.~I. (1977).
\newblock Limit theory for multivariate sample extremes.
\newblock {\em Zeitschrift f{\"u}r Wahrscheinlichkeitstheorie und verwandte
  Gebiete}, 40(4):317--337.

\bibitem[Deng and Qiu, 2021]{deng2021backtesting}
Deng, K. and Qiu, J. (2021).
\newblock Backtesting expected shortfall and beyond.
\newblock {\em Quantitative Finance}, 21(7):1109--1125.

\bibitem[Dhaene et~al., 2003]{VaRsubaddprob}
Dhaene, J., Goovaerts, M.~J., and Kaas, R. (2003).
\newblock {Economic Capital Allocation Derived from Risk Measures}.
\newblock {\em North American Actuarial Journal}, 7(2):44--56.

\bibitem[Dhaene et~al., 2022]{dhaene2022systemic}
Dhaene, J., Laeven, R.~J., and Zhang, Y. (2022).
\newblock Systemic risk: Conditional distortion risk measures.
\newblock {\em Insurance: Mathematics and Economics}, 102:126--145.

\bibitem[Ding, 2016]{condtdist}
Ding, P. (2016).
\newblock {On the Conditional Distribution of the Multivariate t Distribution}.
\newblock {\em The American Statistician}, 70(3):293--295.

\bibitem[Donnelly and Embrechts, 2010]{embrechts_2010}
Donnelly, C. and Embrechts, P. (2010).
\newblock {The Devil is in the Tails: Actuarial Mathematics and the Subprime
  Mortgage Crisis}.
\newblock {\em ASTIN Bulletin}, 40(1):1–33.

\bibitem[Embrechts, 2009]{embrechts2009}
Embrechts, P. (2009).
\newblock {Copulas: A Personal View}.
\newblock {\em Journal of Risk and Insurance}, 76(3):639--650.

\bibitem[Embrechts et~al., 2013]{embrechts2013modelling}
Embrechts, P., Kl{\"u}ppelberg, C., and Mikosch, T. (2013).
\newblock {\em Modelling extremal events: for insurance and finance},
  volume~33.
\newblock Springer Science \& Business Media.

\bibitem[Embrechts et~al., 2009]{embrechts2009multivariate}
Embrechts, P., Lambrigger, D.~D., and W{\"u}thrich, M.~V. (2009).
\newblock Multivariate extremes and the aggregation of dependent risks:
  examples and counter-examples.
\newblock {\em Extremes}, 12(2):107--127.

\bibitem[Embrechts et~al., 2001]{embrechts2001correlation}
Embrechts, P., McNeil, A.~J., and Straumann, D. (2001).
\newblock Correlation and dependence in risk management.
\newblock {\em Risk management: value at risk and beyond’, Cambridge
  University Press, Cambridge}.

\bibitem[Embrechts et~al., 2003]{ModexteventsEmbr}
Embrechts, P., Mikosch, T., and Kluppelberg, C. (2003).
\newblock {\em Modelling extremal events: for insurance and finance}.
\newblock Applications of mathematics 33. Springer, Berlin.

\bibitem[Embrechts et~al., 1999]{embrechts1999extreme}
Embrechts, P., Resnick, S.~I., and Samorodnitsky, G. (1999).
\newblock Extreme value theory as a risk management tool.
\newblock {\em North American Actuarial Journal}, 3(2):30--41.

\bibitem[Engle, 2002]{engle2002dynamic}
Engle, R. (2002).
\newblock Dynamic conditional correlation: A simple class of multivariate
  generalized autoregressive conditional heteroskedasticity models.
\newblock {\em Journal of Business \& Economic Statistics}, 20(3):339--350.

\bibitem[Farmer and Lillo, 2004]{Pl1}
Farmer, J.~D. and Lillo, F. (2004).
\newblock On the origin of power-law tails in price fluctuations.
\newblock {\em Quantitative Finance}, 4(1):C7.

\bibitem[Fissler and Hoga, 2021]{fissler2021backtesting}
Fissler, T. and Hoga, Y. (2021).
\newblock Backtesting systemic risk forecasts using multi-objective
  elicitability.
\newblock {\em arXiv preprint arXiv:2104.10673}.

\bibitem[Freixas et~al., 2015]{sysriskbook}
Freixas, X., Laeven, L., and Peydro, J.-L. (2015).
\newblock {\em Systemic Risk, Crises, and Macroprudential Regulation},
  volume~1.
\newblock The MIT Press, 1 edition.

\bibitem[Freixas and Rochet, 2008]{freixas2008microeconomics}
Freixas, X. and Rochet, J. (2008).
\newblock {\em Microeconomics of Banking, Second Edition}.
\newblock Mit Press. MIT Press.

\bibitem[Gabaix, 2009]{PL3}
Gabaix, X. (2009).
\newblock Power laws in economics and finance.
\newblock {\em Annu. Rev. Econ.}, 1(1):255--294.

\bibitem[Gabaix, 2012]{gabaix2012variable}
Gabaix, X. (2012).
\newblock Variable rare disasters: An exactly solved framework for ten puzzles
  in macro-finance.
\newblock {\em The Quarterly journal of economics}, 127(2):645--700.

\bibitem[Gabaix, 2016]{PL7}
Gabaix, X. (2016).
\newblock Power laws in economics: An introduction.
\newblock {\em Journal of Economic Perspectives}, 30(1):185--206.

\bibitem[Gabaix et~al., 2003]{PLth}
Gabaix, X., Gopikrishnan, P., Plerou, V., and Stanley, H.~E. (2003).
\newblock A theory of power-law distributions in financial market fluctuations.
\newblock {\em Nature}, 423(6937):267--270.

\bibitem[Genest et~al., 2007]{genest2007everything}
Genest, C., Favre, A.-C., et~al. (2007).
\newblock Everything you always wanted to know about copula modeling but were
  afraid to ask.
\newblock {\em Journal of hydrologic engineering}, 12(4):347--368.

\bibitem[Genest et~al., 2009]{Genest2009}
Genest, C., Gendron, M., and Bourdeau-Brien, M. (2009).
\newblock {The Advent of Copulas in Finance}.
\newblock {\em The European Journal of Finance}, 15(7-8):609--618.

\bibitem[Girardi and {Tolga Ergün}, 2013]{GIRARDI20133169}
Girardi, G. and {Tolga Ergün}, A. (2013).
\newblock {Systemic risk measurement: Multivariate GARCH estimation of CoVaR}.
\newblock {\em Journal of Banking \& Finance}, 37(8):3169--3180.

\bibitem[Gnedenko, 1943]{gnedenko1943distribution}
Gnedenko, B. (1943).
\newblock Sur la distribution limite du terme maximum d'une serie aleatoire.
\newblock {\em Annals of mathematics}, pages 423--453.

\bibitem[Gneiting, 2011]{gneiting2011backtest}
Gneiting, T. (2011).
\newblock Making and evaluating point forecasts.
\newblock {\em Journal of the American Statistical Association},
  106(494):746--762.

\bibitem[Gopikrishnan et~al., 1999]{PL5}
Gopikrishnan, P., Plerou, V., Amaral, L. A.~N., Meyer, M., and Stanley, H.~E.
  (1999).
\newblock Scaling of the distribution of fluctuations of financial market
  indices.
\newblock {\em Physical Review E}, 60(5):5305.

\bibitem[Gudendorf and Segers, 2010]{EVcopbook}
Gudendorf, G. and Segers, J. (2010).
\newblock Extreme-value copulas.
\newblock In {\em Copula theory and its applications}, pages 127--145.
  Springer.

\bibitem[Hampel, 1971]{hampel1971general}
Hampel, F.~R. (1971).
\newblock A general qualitative definition of robustness.
\newblock {\em The annals of mathematical statistics}, 42(6):1887--1896.

\bibitem[Hill, 1975]{hill1975simple}
Hill, B.~M. (1975).
\newblock A simple general approach to inference about the tail of a
  distribution.
\newblock {\em The annals of statistics}, pages 1163--1174.

\bibitem[Hofert et~al., 2019]{hofert2019elements}
Hofert, M., Kojadinovic, I., M{\"a}chler, M., and Yan, J. (2019).
\newblock {\em Elements of Copula Modeling with R}.
\newblock Use R! Springer International Publishing.

\bibitem[Hofert et~al., 2022]{Coppackmanual}
Hofert, M., Kojadinovic, I., Maechler, M., and Yan, J. (2022).
\newblock {\em {copula: Multivariate Dependence with Copulas}}.
\newblock R package version 1.1-0.

\bibitem[Hu et~al., 2022]{hu2022tail}
Hu, W., Chen, C., Shi, Y., and Chen, Z. (2022).
\newblock {A Tail Measure With Variable Risk Tolerance: Application in Dynamic
  Portfolio Insurance Strategy}.
\newblock {\em Methodology and Computing in Applied Probability}, pages 1--44.

\bibitem[Jaworski, 2017]{Jaworski+2017+1+19}
Jaworski, P. (2017).
\newblock {On Conditional Value at Risk (CoVaR) for tail-dependent copulas}.
\newblock {\em Dependence Modeling}, 5(1):1--19.

\bibitem[Jessen and Mikosch, 2006]{jessen2006regularly}
Jessen, H.~A. and Mikosch, T. (2006).
\newblock Regularly varying functions.
\newblock {\em Publications de L'institut Mathematique}, 80(94):171--192.

\bibitem[Ji et~al., 2018]{ji2018uncertainties}
Ji, Q., Liu, B.-Y., Nehler, H., and Uddin, G.~S. (2018).
\newblock Uncertainties and extreme risk spillover in the energy markets: A
  time-varying copula-based covar approach.
\newblock {\em Energy Economics}, 76:115--126.

\bibitem[Jondeau and Rockinger, 2006]{JONDEAU2006827}
Jondeau, E. and Rockinger, M. (2006).
\newblock {The Copula-GARCH model of conditional dependencies: An international
  stock market application}.
\newblock {\em Journal of International Money and Finance}, 25(5):827--853.

\bibitem[Karimalis and Nomikos, 2018]{CopulaGarchCovar}
Karimalis, E.~N. and Nomikos, N.~K. (2018).
\newblock {Measuring systemic risk in the European banking sector: a copula
  CoVaR approach}.
\newblock {\em The European Journal of Finance}, 24(11):944--975.

\bibitem[Keilbar and Wang, 2022]{NetwNN}
Keilbar, G. and Wang, W. (2022).
\newblock Modelling systemic risk using neural network quantile regression.
\newblock {\em Empirical Economics}, 62(1):93--118.

\bibitem[Kelly, 2014]{kelly2014dynamic}
Kelly, B. (2014).
\newblock The dynamic power law model.
\newblock {\em Extremes}, 17(4):557--583.

\bibitem[Kelly and Jiang, 2014]{kelly2014tail}
Kelly, B. and Jiang, H. (2014).
\newblock Tail risk and asset prices.
\newblock {\em The Review of Financial Studies}, 27(10):2841--2871.

\bibitem[Kuester et~al., 2005]{Condvarcompari}
Kuester, K., Mittnik, S., and Paolella, M.~S. (2005).
\newblock {Value-at-Risk Prediction: A Comparison of Alternative Strategies}.
\newblock {\em Journal of Financial Econometrics}, 4(1):53--89.

\bibitem[Landsman and Valdez, 2003]{EllipticalDists}
Landsman, Z.~M. and Valdez, E.~A. (2003).
\newblock {Tail Conditional Expectations for Elliptical Distributions}.
\newblock {\em North American Actuarial Journal}, 7(4):55--71.

\bibitem[Ledford and Tawn, 1996]{Evcopestitest}
Ledford, A.~W. and Tawn, J.~A. (1996).
\newblock Statistics for near independence in multivariate extreme values.
\newblock {\em Biometrika}, 83(1):169--187.

\bibitem[Longin and Solnik, 2001]{EVcopfinapp}
Longin, F. and Solnik, B. (2001).
\newblock Extreme correlation of international equity markets.
\newblock {\em The journal of finance}, 56(2):649--676.

\bibitem[Mainik and Schaanning, 2014]{mainik2014dependence}
Mainik, G. and Schaanning, E. (2014).
\newblock {On dependence consistency of CoVaR and some other systemic risk
  measures}.
\newblock {\em Statistics \& Risk Modeling}, 31(1):49--77.

\bibitem[Malevergne et~al., 2005]{PL4}
Malevergne, Y., Pisarenko, V., and Sornette, D. (2005).
\newblock Empirical distributions of stock returns: between the stretched
  exponential and the power law?
\newblock {\em Quantitative Finance}, 5(4):379--401.

\bibitem[McNeil and Frey, 2000]{Garchevt}
McNeil, A.~J. and Frey, R. (2000).
\newblock Estimation of tail-related risk measures for heteroscedastic
  financial time series: an extreme value approach.
\newblock {\em Journal of Empirical Finance}, 7(3):271--300.
\newblock Special issue on Risk Management.

\bibitem[McNeil et~al., 2015]{mcneil2015quantitative}
McNeil, A.~J., Frey, R., and Embrechts, P. (2015).
\newblock {\em Quantitative risk management: concepts, techniques and
  tools-revised edition}.
\newblock Princeton university press.

\bibitem[{Microsoft} and Weston, 2022]{foreach}
{Microsoft} and Weston, S. (2022).
\newblock {\em foreach: Provides Foreach Looping Construct}.
\newblock R package version 1.5.2.

\bibitem[Nadarajah et~al., 2014]{ESestgen}
Nadarajah, S., Zhang, B., and Chan, S. (2014).
\newblock Estimation methods for expected shortfall.
\newblock {\em Quantitative Finance}, 14(2):271--291.

\bibitem[Nelsen, 2007]{nelsen2007introduction}
Nelsen, R. (2007).
\newblock {\em An Introduction to Copulas}.
\newblock Springer Series in Statistics. Springer New York.

\bibitem[Nolde and Zhou, 2021]{nolde2021extreme}
Nolde, N. and Zhou, C. (2021).
\newblock Extreme value analysis for financial risk management.
\newblock {\em Annual Review of Statistics and Its Application}, 8:217--240.

\bibitem[{Norton} et~al., 2018]{ESexactT}
{Norton}, M., {Khokhlov}, V., and {Uryasev}, S. (2018).
\newblock {Calculating CVaR and bPOE for Common Probability Distributions With
  Application to Portfolio Optimization and Density Estimation}.
\newblock {\em arXiv e-prints}, page arXiv:1811.11301.

\bibitem[NYU, 2022]{vlab}
NYU (2022).
\newblock {V-Lab}.
\newblock \url{https://vlab.stern.nyu.edu/}.

\bibitem[Pickands, 1989]{EVcop2}
Pickands, J. (1989).
\newblock Multivariate negative exponential and extreme value distributions.
\newblock In {\em Extreme Value Theory}, pages 262--274. Springer.

\bibitem[Pickands~III, 1975]{pickands1975statistical}
Pickands~III, J. (1975).
\newblock Statistical inference using extreme order statistics.
\newblock {\em the Annals of Statistics}, pages 119--131.

\bibitem[Plerou et~al., 1999]{PL6}
Plerou, V., Gopikrishnan, P., Amaral, L. A.~N., Meyer, M., and Stanley, H.~E.
  (1999).
\newblock Scaling of the distribution of price fluctuations of individual
  companies.
\newblock {\em Physical review e}, 60(6):6519.

\bibitem[Plerou et~al., 2004]{PL2}
Plerou, V., Gopikrishnan, P., Gabaix, X., and Stanley, H.~E. (2004).
\newblock On the origin of power-law fluctuations in stock prices.
\newblock {\em Quantitative Finance}, 4(1):C11.

\bibitem[{R Core Team}, 2021]{Rsoftware}
{R Core Team} (2021).
\newblock {\em R: A Language and Environment for Statistical Computing}.
\newblock R Foundation for Statistical Computing, Vienna, Austria.

\bibitem[Reboredo, 2013]{REBOREDO20132665}
Reboredo, J.~C. (2013).
\newblock Is gold a safe haven or a hedge for the us dollar? implications for
  risk management.
\newblock {\em Journal of Banking and Finance}, 37(8):2665--2676.

\bibitem[Reboredo and Ugolini, 2015]{REBOREDO2015214}
Reboredo, J.~C. and Ugolini, A. (2015).
\newblock {Systemic risk in European sovereign debt markets: A CoVaR-copula
  approach}.
\newblock {\em Journal of International Money and Finance}, 51:214--244.

\bibitem[Ryan and Ulrich, 2022]{xts}
Ryan, J.~A. and Ulrich, J.~M. (2022).
\newblock {\em xts: eXtensible Time Series}.
\newblock R package version 0.12.2.

\bibitem[Segers et~al., 2017]{SEGERS201735}
Segers, J., Sibuya, M., and Tsukahara, H. (2017).
\newblock The empirical beta copula.
\newblock {\em Journal of Multivariate Analysis}, 155:35--51.

\bibitem[Sklar, 1959]{sklar1959fonctions}
Sklar, A. (1959).
\newblock Fonctions de répartition à n dimensions et leurs marges.
\newblock {\em Publ. inst. statist. univ. Paris}, 8:229--231.

\bibitem[Song and Fang, 2022]{song2022temperature}
Song, X. and Fang, T. (2022).
\newblock Temperature shocks and bank systemic risk: Evidence from china.
\newblock {\em Finance Research Letters}, page 103447.

\bibitem[Sordo et~al., 2018]{SORDO2018105}
Sordo, M., Bello, A., and Suárez-Llorens, A. (2018).
\newblock {Stochastic orders and co-risk measures under positive dependence}.
\newblock {\em Insurance: Mathematics and Economics}, 78:105--113.
\newblock Longevity risk and capital markets: The 2015–16 update.

\bibitem[Tawn, 1988]{Evcopesti1}
Tawn, J.~A. (1988).
\newblock Bivariate extreme value theory: models and estimation.
\newblock {\em Biometrika}, 75(3):397--415.

\bibitem[Torri et~al., 2021]{TORRI2021104125}
Torri, G., Giacometti, R., and Tichý, T. (2021).
\newblock {Network tail risk estimation in the European banking system}.
\newblock {\em Journal of Economic Dynamics and Control}, 127:104125.

\bibitem[Wei and Simko, 2021]{Rcorrplot2021}
Wei, T. and Simko, V. (2021).
\newblock {\em R package 'corrplot': Visualization of a Correlation Matrix}.
\newblock (Version 0.92).

\bibitem[Wickham et~al., 2019]{RTidyverse}
Wickham, H., Averick, M., Bryan, J., Chang, W., McGowan, L.~D., François, R.,
  Grolemund, G., Hayes, A., Henry, L., Hester, J., Kuhn, M., Pedersen, T.~L.,
  Miller, E., Bache, S.~M., Müller, K., Ooms, J., Robinson, D., Seidel, D.~P.,
  Spinu, V., Takahashi, K., Vaughan, D., Wilke, C., Woo, K., and Yutani, H.
  (2019).
\newblock Welcome to the {tidyverse}.
\newblock {\em Journal of Open Source Software}, 4(43):1686.

\bibitem[Wickham et~al., 2021]{Rdplyr}
Wickham, H., François, R., Henry, L., and Müller, K. (2021).
\newblock {\em dplyr: A Grammar of Data Manipulation}.
\newblock R package version 1.0.7.

\bibitem[Wickham and Miller, 2021]{Rhaven}
Wickham, H. and Miller, E. (2021).
\newblock {\em haven: Import and Export 'SPSS', 'Stata' and 'SAS' Files}.
\newblock R package version 2.4.3.

\bibitem[Yamai and Yoshiba, 2002]{yamai2002comparative}
Yamai, Y. and Yoshiba, T. (2002).
\newblock {{Comparative Analyses of Expected Shortfall and Value-at-Risk (3):
  Their Validity under Market Stress}}.
\newblock {\em Monetary and Economic Studies, Bank of Japan}, 20(3):181--237.

\bibitem[Zeileis and Grothendieck, 2005]{Rzoo}
Zeileis, A. and Grothendieck, G. (2005).
\newblock zoo: S3 infrastructure for regular and irregular time series.
\newblock {\em Journal of Statistical Software}, 14(6):1--27.

\bibitem[Zelenyuk and Faff, 2022]{zelenyuk2022effects}
Zelenyuk, N. and Faff, R. (2022).
\newblock Effects of incentive pay on systemic risk: evidence from ceo
  compensation and covar.
\newblock {\em Empirical Economics}, pages 1--23.

\end{thebibliography}
\newpage
\section{Appendix}\label{Sec:appendix}
\subsection{Theoretical Analysis: $\Delta$-CoES under normality conditional on $X=\text{VaR}_q(X)$}\label{App:Theor:norm}
Just as \cite{AdrianBrunnCovar} we assume that the losses of the financial system and an institution $i$ follow a bivariate normal distribution. Hence,
\begin{center}
   $\begin{pmatrix}X^{\text{sys}}_t\\X^i_t \end{pmatrix}\sim\mathcal{N}\left(\begin{pmatrix}0\\0  \end{pmatrix}, \boldsymbol{\Sigma}  \right)$
\end{center}
with
\begin{center}
    $\boldsymbol{\Sigma}=\begin{pmatrix}(\sigma_t^\text{sys})^2 &\sigma_t^\text{sys}\sigma_t^i\rho_t\\\sigma_t^\text{sys}\sigma_t^i\rho_t &(\sigma_t^i)^2\end{pmatrix}$.
\end{center}
Assuming a linear relationship between the two variables it follows that: 
\begin{center}
    $X_t^{\text{sys}}\mid X_t^i\sim\mathcal{N}\left(\frac{X_t^{i} \sigma_t^\text{sys}\rho_t}{\sigma_t^i}, (1-\rho^2_t) ({\sigma_t^\text{sys}})^2  \right)$.
\end{center}
From the results of \cite{AdrianBrunnCovar} it follows that:
\begin{equation}
    \CoV^{=,t}_{\alpha,\beta}=\Phi^{-1}(\beta)\sqrt{1-\rho_t^2}\sigma_t^\text{sys}+\Phi^{-1}(\alpha)\sigma_t^{\text{sys}}\rho_t
\end{equation}
\noindent with $\beta$ the level of the CoVaR and $\alpha$ the level of the VaR of $X^i_t$. If $\alpha=\beta$ Then  
\begin{equation}
    \DCov^{=,\text{med},t}_{\beta}=\Phi^{-1}(\beta)\sigma_t^{\text{sys}}\rho_t,
\end{equation}
\noindent Similarly, using the ES formula from \cite{ESexactT} it follows that:
\begin{equation}
    \DCoES^{=,t}_{\alpha,\beta}=\Phi^{-1}(\beta)\sqrt{1-\rho_t^2}\sigma_t^\text{sys}\frac{\phi(\Phi^{-1}(\beta)}{\beta}+\Phi^{-1}(\alpha)\sigma_t^{\text{sys}}\rho_t.
\end{equation}
\noindent If $\alpha=\beta$ then
\begin{equation}
    \DCoES^{=,\text{med},t}_\beta=\Phi^{-1}(\beta)\sigma_t^{\text{sys}}\rho_t.
\end{equation}
\noindent Therefore, under normality both systemic risk measures are equivalent. This is not surprising as the normal distribution is fully determined by its mean and variance and since we condition on an event of probability zero ($X^i_t=$VaR$(X^i_t)$) the tail of $X^i_t$ beyond this level is not taken into account. This leads to straightforward expressions for the CoVaR and CoES which are very similar. As the term that does differ drops out when computing the $\Delta$ measures we end up with equivalent results for both.   
\subsection{Theroretical Analysis: $\Delta$-CoVaR \ $\Delta$-CoES under a t-distribution conditional on $X=\text{VaR}_q(X)$}\label{App:theor:tgen}
Now we assume the losses of the system and institution $i$ follow a bivariate generalized t-distribution with degrees of freedom $\nu$. Hence,
\begin{center}
    \begin{center}
    $\begin{pmatrix}X^{\text{sys}}_t\\X^i_t \end{pmatrix}\sim T\left(\begin{pmatrix}0\\0  \end{pmatrix}, \boldsymbol{\Sigma}  ,\nu\right)$
\end{center}
\end{center}
with
\begin{center}
    $\boldsymbol{\Sigma}=\begin{pmatrix}(\sigma_t^\text{sys})^2&\sigma_t^\text{sys}\sigma_t^i\rho_t\\\sigma_t^\text{sys}\sigma_t^i\rho_t&(\sigma_t^i)^2\end{pmatrix}$.
\end{center}
\noindent Then applying the result of \cite{condtdist} we have that:
\begin{center}
    $X_t^{\text{sys}}\mid X_t^i\sim T\left(\frac{X_t^{i} \sigma_t^\text{sys}\rho_t}{\sigma_t^i}, \frac{\nu+d_1}{\nu+1}(1-\rho^2_t) (\sigma_t^\text{sys})^2,\nu +1  \right)$
\end{center}
with $d_1=\left(\frac{X^i_t}{{\sigma_t^i}}\right)^2$. Therefore, the conditional distribution is also a generalized t-distribution. Now using similar reasoning as in \cite{AdrianBrunnCovar} we obtain that
\begin{center}
    $S=\left(\frac{X_t^{\text{sys}}-X_t^{i} \sigma_t^\text{sys}\rho_t/\sigma_t^i}{\sqrt{\frac{\nu+d_1}{\nu+1}(1-\rho^2_t) (\sigma_t^\text{sys})^2}}\right)\sim T(0,1,\nu+1)$.
\end{center}
\noindent Hence, $\Expect[S]=0,\var(S)=\frac{\nu+1}{\nu-1}$ as $S$ follows a standardized t-distribution. The VaR of $i$ is then $\text{VaR}^{i}_{q,t}=\sigma^i_t\cdot T_\nu^{-1}(q)$ where $T_\nu^{-1}(q)$ is the inverse cdf of the standardized t-distribution with $\nu$ degrees of freedom. Setting $X^i_t=\text{VaR}^{i}_{q,t}$, applying the definition of the CoVar and solving for the CoVaR we obtain
\begin{equation}
    \CoV^{=,t}_{\alpha,\beta}= T_{\nu+1}^{-1}(\beta)\sqrt{\frac{\nu+d_1}{\nu+1}(1-\rho^2_t)}{\sigma_t^\text{sys}}+T_{\nu}^{-1}(\alpha)\sigma_t^\text{sys}\rho_t.
\end{equation}
\noindent With $d_1=(T^{-1}_\nu(\beta))^2$.We can check the result by taking $\nu\xrightarrow{}\infty$. As expected the result is equivalent to the CoVaR under normality. if $\alpha=\beta$. Then it follows that:
\begin{equation}
   \DCov^{=,\text{med},t}_{\beta}= T_{\nu}^{-1}(\beta)\sigma_t^{\text{sys}}\rho_t.
\end{equation}
\noindent Using the ES formula from \cite{ESexactT} we obtain
\begin{equation}
    \CoES^{=,t}_{\alpha,\beta}=\tau_{\nu+1}(T_{\nu+1}^{-1}(\beta))\left(\frac{\nu+1+T_{\nu+1}^{-1}(\beta)^2}{\nu(1-\beta)}\right)\sqrt{\frac{\nu+d_1}{\nu+1}(1-\rho^2_t)}{\sigma_t^\text{sys}}+T_{\nu}^{-1}(\alpha)\sigma_t^{\text{sys}}\rho_t
\end{equation}
where $\tau_\nu(x)$ is the pdf of the standardized t-distribution. Again, taking $\nu\xrightarrow{}\infty$ results in the CoES expression under normality. As before if $\alpha=\beta$ we then get that
\begin{equation}
    \DCoES^{=,\text{med},t}_{\beta}= T_{\nu}^{-1}(\beta)\sigma_t^{\text{sys}}\rho_t.
\end{equation}
\noindent This result is interesting because with $\nu$ the kurtosis can be made arbitrarily large. It must be noted that all these results are only valid for $\nu+1>2$. While this excludes a multivariate Cauchy distribution ($\nu=1$) it encompasses distributions whose excess kurtosis can be made arbitrarily large ( for a standardized t-distribution this is $\frac{6}{\nu-4}$ for $\nu>4$ otherwise the kurtosis does not exist). Also, through $T_\nu$ some of the power-law tail of the t-distribution is captured. It must be noted that all these results hinge on a linear relationship (or at least approximately linear) between the losses of the financial system and an institution $i$.  This result was not obtained in earlier work such as \cite{mainik2014dependence} and again shows how conditioning on $X_t^i=\text{VaR}(X)$ leads to risk measures which fail to capture the tail.  We conjecture this equivalence result can be further generalized to the family of elliptical distributions (except the Cauchy) as at least for the ES-based measures the structure seems to be known and similar across this whole family \cite{EllipticalDists}. Because elliptical distributions are popular in (joint) risk modelling \cite{mcneil2015quantitative} such a result would call even more into question the use of the conditioning on $X_t^i=\text{VaR}(X)$.
\subsection{Proof of ES and MES representation}\label{App:proofESMESrep}
\begin{proof}The claim is that: 
\begin{equation*}
    \CoES_{\alpha,\beta}(Y\mid X)=\ES_\omega(Y)=\frac{1}{1-\omega}\int_{\omega}^1\VaR_q(Y) dq.
\end{equation*}
\noindent which reduces to proving that:
\begin{equation*}
    \frac{1}{1-\beta}\int_{\beta}^1\CoV_{\alpha,p}(Y\mid X)dp=\frac{1}{1-\omega}\int_{\omega}^1 \VaR_q(Y)dq.
\end{equation*}
We can write this as, see definition 2.2 in \cite{mainik2014dependence}:
\begin{equation*}
    \frac{1}{1-\beta}\int_{\beta}^1 F^{-1}_{Y\mid X\geq \VaR_\alpha(X)}(p)dp=\frac{1}{1-\omega}\int_{\omega}^1  F^{-1}_Y(q) dq.
\end{equation*}
We will now apply a change of variables to the left-hand side where $q=\omega(\alpha,p,C)$ with $\omega(\alpha,p,C)$ the largest solution to $\bar{C}(\alpha,\omega)=(1-p)(1-\alpha)$. Then, the result of \cite{BERNARDI20178} shows that $F^{-1}_{Y\mid X\geq \VaR_\alpha(X)}(p)=F^{-1}_Y(q)$. Looking at the range of $p$ we get for any $p\in[\beta,1]$ that:
\begin{equation*}
    1-\alpha-\omega(\alpha,p,C)+C(\alpha,\omega(\alpha,p,C))=(1-p)(1-\alpha).
\end{equation*}
By definition we get that if $p=\beta$ then $q=\omega(\alpha,\beta,C)$. Now if $p=1$ we get that:
\begin{equation*}
    1-\alpha-\omega(\alpha,1,C)+C(\alpha,\omega(\alpha,1,C))=0.
\end{equation*}
Therefore, $\omega(\alpha,1,C)$ is the largest solution to $C(\alpha,\omega(\alpha,1,C))=\alpha+\omega(\alpha,1,C)-1$. According to the basic properties of copulas $\omega(\alpha,1,C)=1$ because then $C(\alpha,1)=\alpha$ \cite{nelsen2007introduction}. Therefore $q=1$. Because if $p=\beta$ then $q=\omega(\alpha,\beta,C)$ (for convenience shortened to $\omega$) and the definition of the $\ES$ $\frac{1}{1-\omega}$ is obtained.  
\end{proof}
\begin{proof}
For the MES this reduces to setting $\beta=0$ in the proof above. However, here one must be aware of possible multiple solutions which when the rule to take the largest solution shows its utility. 
\end{proof}
\subsection{Proofs of coherence, independence and invariance}
We must show that for a bivariate random vector $(X,Y)$ with copula $uv\leq C(u,v)\leq \min(u,v)$ for all $(u,v)\in(0,1)^2$ the following holds:
\begin{equation*}
    \DCov_{\alpha,\beta}(Y\mid X)=0 \iff X,Y \text{ are independent,}
\end{equation*}
\begin{equation*}
    \DCoES_{\alpha,\beta}(Y\mid X)=0 \iff X,Y \text{ are independent,}
\end{equation*}
\begin{equation*}
    \DCov_{\alpha,\beta}(Y\mid X)\text{ is maximal}\iff X,Y \text{ are comonotonic and } 
\end{equation*}
\begin{equation*}
    \DCoES_{\alpha,\beta}(Y\mid X)\text{ is maximal}\iff X,Y \text{ are comonotonic. } 
\end{equation*}
\begin{proof}
First we prove the $\impliedby$ of all equivalences. This is rather easy as for the first two equivalences by independence we have $\omega=\beta$ and the implication follows. For the third and fourth equivalences using the fact that the $\DCoES$ and $\DCov$ are dependence consistent under these conditions \cite{SORDO2018105,dhaene2022systemic} we obtain that under any copula $C(u,v)\leq \min(u,v)$ we get that $\omega\leq \alpha+\beta-\alpha\beta $ and it follows that $\CoV_{\alpha,\beta}(Y\mid X)=\VaR_\omega(Y)\leq \VaR_{\alpha+\beta-\alpha\beta}(Y)$ and $\CoES_{\alpha,\beta}(Y\mid X)=\ES_\omega(Y)\leq \ES_{\alpha+\beta-\alpha\beta}(Y)$. By the definition of the $\DCoES$ and $\DCov$ it then also follows this upper bound holds for the $\DCoES$ and $\DCov$. \\
To prove the $\implies$ for all equivalences by taking the contrapositive one gets implications that are easy to prove. For example, take $X,Y$ are not comonotonic $\implies\DCoES_{\alpha,\beta}(Y\mid X)\text{ is not maximal}$. We know that $\omega< \alpha+\beta-\alpha\beta$ and hence that $\CoES_{\alpha,\beta}(Y\mid X)=\ES_\omega(Y)< \ES_{\alpha+\beta-\alpha\beta}(Y)$ from which it follows that $\DCoES$ does not attain its upper bound. As these contrapositives are true the original implications are true.
\end{proof}
\subsection{Proofs of tail sensitivity results}\label{App:prooftailsens}
Proposition \ref{prop:tailsensuni}:
\begin{proof}
First, we use a result in \cite{GPDtrick} to obtain an expression of $F_(y)$ from $F_u(y)$. This allows for $\VaR$ and $\ES$ computation for quantiles of $Y$ past that of the threshold $\gamma$. This result states that:
\begin{equation*}
    F(y)=(1-F(u))F_u(y)+F(u).
\end{equation*}
By definition of the threshold $u$ one obtains $F(u)=\gamma$. Because the GPD is an approximation for finite $u$:
\begin{equation*}
    F(y)\approx(1-\gamma)\left(1-\left(1+\xi\frac{y-\VaR_\gamma(Y)}{s}\right)^{-1/\xi}\right)+\gamma.
\end{equation*}
\noindent Since it is assumed that $u$ is sufficiently high for the GPD approximation to be accurate in the following expressions equalities are used. Setting $F(y)=\alpha$ and $y=\VaR_\alpha(Y)$ for some $1>\alpha\geq\gamma$ and rearranging terms we obtain:
\begin{equation*}
    \frac{1-\alpha}{1-\gamma}=\left(1+\xi\frac{\VaR_\alpha(Y)-\VaR_\gamma(Y)}{s}\right)^{-1/\xi}.
\end{equation*}
\noindent Now, solving for $\VaR_\alpha(Y)$ we obtain an expression for the value-at-risk of $Y$ with GPD tails beyond a given threshold.
\begin{equation*}
    \VaR_\alpha(Y)=\VaR_\gamma(Y)+s\left(\frac{\left(\frac{1-\alpha}{1-\gamma}\right)^{-\xi}-1}{\xi}\right).
\end{equation*}
\noindent If $\alpha=\gamma$ the expression simply reduces to $\VaR_\gamma(Y)$. This expression is nothing more than the normal expression for the $\VaR$ of a GPD but now adjusted for computing levels past the threshold. Because of this fact the expression from the $\ES$ follows simply from the expression of the $\ES$ of the GPD. Therefore,
\begin{equation*}
    \ES_\alpha(Y)=\VaR_\gamma(Y)+s\left(\frac{\left(\frac{1-\alpha}{1-\gamma}\right)^{-\xi}}{1-\xi}+\frac{\left(\frac{1-\alpha}{1-\gamma}\right)^{-\xi}-1}{\xi}\right).
\end{equation*}
If $\alpha=\gamma$ the expression reduces to $\VaR_\gamma(Y)+s/(1-\xi)$. Both the $\VaR$ and $\ES$ expressions hold for any quantile $1>\alpha\geq\gamma$. Now, using the representation results for $\CoV,\CoES,\DCov$ and $\DCoES$ results in the claims of proposition \ref{prop:tailsensuni}.
\end{proof}
\noindent For clarity, the relevant expressions are provided below:
\begin{equation*}
    \CoV_{\alpha,\beta}(Y\mid X)=\VaR_\omega(Y)=\VaR_\gamma(Y)+s\left(\frac{\left(\frac{1-\omega}{1-\gamma}\right)^{-\xi}-1}{\xi}\right),
\end{equation*}
\begin{equation*}
    \DCov_{\alpha,\beta}(Y\mid X)=s\left(\frac{\left(\frac{1-\omega}{1-\gamma}\right)^{-\xi}-\left(\frac{1-\beta}{1-\gamma}\right)^{-\xi}}{\xi}\right)
\end{equation*}
\begin{equation*}
    \CoES_{\alpha,\beta}(Y\mid X)=\ES_\omega(Y)=\VaR_\gamma(Y)+s\left(\frac{\left(\frac{1-\omega}{1-\gamma}\right)^{-\xi}}{1-\xi}+ \frac{\left(\frac{1-\omega}{1-\gamma}\right)^{-\xi}-1}{\xi} \right),
\end{equation*}
\begin{equation*}
    \DCoES_{\alpha,\beta}(Y\mid X)=s\left(\frac{\left(\frac{1-\omega}{1-\gamma}\right)^{-\xi}-\left(\frac{1-\beta}{1-\gamma}\right)^{-\xi}}{\xi(1-\xi)}\right)
\end{equation*}
Proposition \ref{prop:tailsensbiv}:
\begin{proof}
Proving the bivariate case reduces to applying the EV copula result from \cite{EVcop1,EVcop2} with a Gumbel copula and then applying the definition of $\omega(\alpha,\beta,C_\theta)$ with $C_\theta$ the Gumbel copula to obtain the result. The results for the risk measures already follow from proposition \ref{prop:tailsensuni}. 
\end{proof}
\subsection{Robustness proofs}\label{App:proofrob}
Proposition \ref{prop:robcoes}: \begin{proof}
By adapting the results of proposition 4.2 and corollary 4.4  \cite{ContRobrisk} to the P/L setting in this paper and applying the $\CoV/\CoES$ representation results the result of the proposition is obtained.  
\end{proof}
Proposition \ref{prop:robdcoes}\begin{proof}
By applying the result of proposition \ref{prop:robcoes} and the definition of the $\DCoES$ the result of the proposition is obtained.
\end{proof}
\subsection{List of institutions}\label{Ssec:inslist}
\begin{table}[H]
    \centering
    \begin{tabular}{|c|c|c|}\hline
         Institution Ticker& Institution name& Classification\\ \hline  
         BAC& Bank of America&Depositories \\
         BBT& BB\& T & Depositories \\
         BK& Bank of New York Mellon & Depositories\\
         C& Citigroup & Depositories \\
         CBH& Commerce Bancorp Inc & Depositories \\
         CMA& Comerica Inc & Depositories \\
         HBAN& Huntingdon Bancshares Inc& Depositories  \\
         HCBK& Hudson City Bancorp Inc& Depositories\\
         JPM& JP Morgan Chase& Depositories\\
         KEY& Keycorp New & Depositories\\
         MI& Marshall Isley& Depositories\\
         MTB& M\&T Bank Corp& Depositories\\
         NCC& National City Corp& Depositories\\
         NTRS& Northern Trust Corp& Depositories\\
         NYB& New York Community Bankcorp& Depositories\\
         PBCT& People United Financial& Depositories\\
         PNC& PNC Financial Services& Depositories\\
         RF& Regions Financials& Depositories\\
         SNV& Synovus Financial Corp& Depositories\\
         SOV& Sovereign Bancorp& Depositories\\
         STI& Suntrust Banks Inc& Depositories\\
         STT& State Street Corp& Depositories\\
         UB& Unionbancal Corp& Depositories\\
         USB& US Bancorp Del& Depositories\\
         WB& Wachovia& Depositories\\
         WFC& Wells Fargo& Depositories\\
         WM& Washington Mutual& Depositories\\
         ZION& Zions Bancorp& Depositories\\\hline
         ACAS& American capital Strategies& Others\\
         AMTD& Ameritrade Holding& Others\\
         AXP& American Express& Others\\
         BEN& Franklin Resources Inc& Others\\
         BLK& Blackrock Inc&Others\\
         COF& Capital One Financial&Others\\
         EV& Eaton Vance Corp& Others\\
         FNM& Federal National Mortgage Assn& Others\\
         FRE& Federal Home Loan Mortgage&Others\\
         JNS& Janus Cap Group Inc&Others\\
         LM& Legg Mason Inc&Others\\
         SEIC& Sei Investments Company&Others\\
         SLM& SLM Corp&Others\\
         CME&CME Group Inc&Others\\
         ICE& Intercontinental Exchange Inc&Others\\
         NDAQ& NASDAQ Inc&Others\\\hline
         
          \end{tabular}
    \caption{Table of financial institutions used for estimating the risk measures Part 1.}
    \label{tab:instab1}
\end{table}       
   \begin{table}[H]
   \centering
   \begin{tabular}{|c|c|c|}\hline
  Institution Ticker& Institution Name & Classification\\\hline
  AFL& AFLA Inc&Insurance\\
         AIG& American International Group&Insurance\\
         ALL& Allstate Corp&Insurance\\
         AON& AON Corp&Insurance\\
         CB&Chubb Corp&Insurance\\
         CFC&Countrywide Financial Corp&Insurance\\
         CINF&Cincinnati Financial Corp&Insurance\\
         CNA& Can Financial Corp& Insurance\\
         HIG& Hartford Financial Svcs Group&Insurance\\
         HUM&Humana Inc&Insurance\\
         L&Loews Corp&Insurance\\
         LNC&Lincoln national Corp&Insurance\\
         MBI& MBIA Inc&Insurance\\
         MET& Metlife Inc& Insurance\\
         MMC& MArsh and Mclennan Cos Inc&Insurance\\
         PGR& Progressive Corp OH&Insurance\\
         SAF& Safeco Corp&Insurance\\
         TMK&Torchmark Corp&Insurance\\
         TRV& Travelers Companies Inc&Insurance\\
         UNH& Untied Health Group&Insurance\\
         UNM& Unum Group&Insurance\\\hline
         BSC& Bear Stearns&Broker-Dealers\\
         ETFC& E-Trade Financial&Broker-Dealers\\
         GS&Goldman Sachs&Broker-Dealers\\
         LEH&Lehman Brothers&Broker-Dealers\\
         MER&Merrill Lynch& Broker-Dealers\\
         MS& Morgan Stanley&Broker-Dealers\\
         SCHW&Charles Schawb Group& Broker-Dealers\\
         TROW& T Rowe Price&Broker-Dealers\\\hline
     \end{tabular}  
     \caption{Table of financial institutions used for estimating the risk measures Part 2.}
     \label{tab:instab2}
\end{table}
\subsection{Data Filtering procedure}\label{Ssec:datafilt}
The table below contains the filtering procedures and all variables of the dataset.
\begin{table}[H]\label{Tab:datfilt}

    \begin{tabular}{|l|l|l|}\hline
         Variable& Filter applied on WRDS& Reason  \\\hline
         PERMNO&  Only PERMNOs from the Brunnermeier dataset & Consistency with said dataset\\\hline 
         SIC& include 6000-6800 exclude all the rest & Same as Brunnermeier\\\hline
         Share code& include only if $<200$& To exclude ADRs, SDIs, REITs etc.\\\hline
         returns&$>-66$& Excludes missing values. \\\hline
         Price (PRC)&$>0$& To exclude zero and negative prices\\\hline
         Delist&$<200$&To exclude inactive firms\\\hline
         Ticker&No restriction& None needed\\\hline
         date& 31-12-1970 to 31-12-2020& Extend time span to maximum\\\hline
         firm name& No restriction& None needed\\\hline
         PERMCO&No restriction&None needed\\\hline
         Shares (SHROUT)&No restirction&None needed\\\hline
    \end{tabular}
    \caption{All variables obtained from CRSP and filters/restrictions.}
    \label{tab:my_label}
\end{table}
\noindent As stated in Section \ref{Sec:data} the returns data were further filtered to include only firms that have at least 260 weeks of returns data. Since the assumption is that trading weeks consist of 5 days this implies including firms that have at least 1300 trading days of returns. This resulted in the sample size of firms decreasing to 1564 from 1688. None of the variables have been winsorized as this would eliminate the extreme events we are interested in. Also, due to discrepancies noticed between our dataset and that of Brunnermeier we also downloaded a version from CRSP without any restrictions and compared the SIC codes between the two datasets. From this it became apparent that the Brunnermeier datset contains firms whose SIC codes are not between 6000 and 6800. Before the filter on the amount of returns days this difference amounts to 135 firms while after the filter it has increased to 259 firms.\newline

\noindent As noted in Section \ref{Sec:data} Mastercard and Visa have been excluded. This applies to Paypal as well as its registered under the same SIC code (7389) as Visa and Mastercard. Hence, the three firms are excluded from the financial sector index. The exclusion could pose issues considering how representative said index is as the market capitalisation of all three firms are among the top 50 of the S\&P 500 in terms of market capitalisation. However, as these firms mainly provide payment processing their systemic risk profile might be different from more traditional financial firms which tend to have interconnected claims against each other. We note though that according to \cite{freixas2008microeconomics} large value payment systems are given as one of the 4 sources of financial contagion.    
\subsection{Supplementary Tables and Figures}\label{App:supps}
\begin{table}[H]
\centering
\begin{tabular}{ccccc}
  \hline
 $n$& $\DCov$ & $\DCoES$ & $\omega$ & $\xi$ \\ 
  \hline

  500 & 4.195 & 19.753 & 0.000000146 & 0.181 \\ 
  1000 & 2.245 & 14.957 & 0.000000041 & 0.082 \\ 
  2000 & 1.240 & 9.024 & 0.000000015 & 0.045 \\ 
  5000 & 0.513 & 3.936 & 0.000000005 & 0.020 \\ 
  10000 & 0.258 & 1.948 & 0.000000002 & 0.011 \\ 
  20000 & 0.131 & 0.991 & 0.000000001 & 0.006 \\ 
   \hline
   
\end{tabular}
\caption{MSE of the estimates from the simulation study}
   \label{tab:mse}
\end{table}
\begin{figure}[H]
    \centering
    \includegraphics[width=1.2\textwidth]{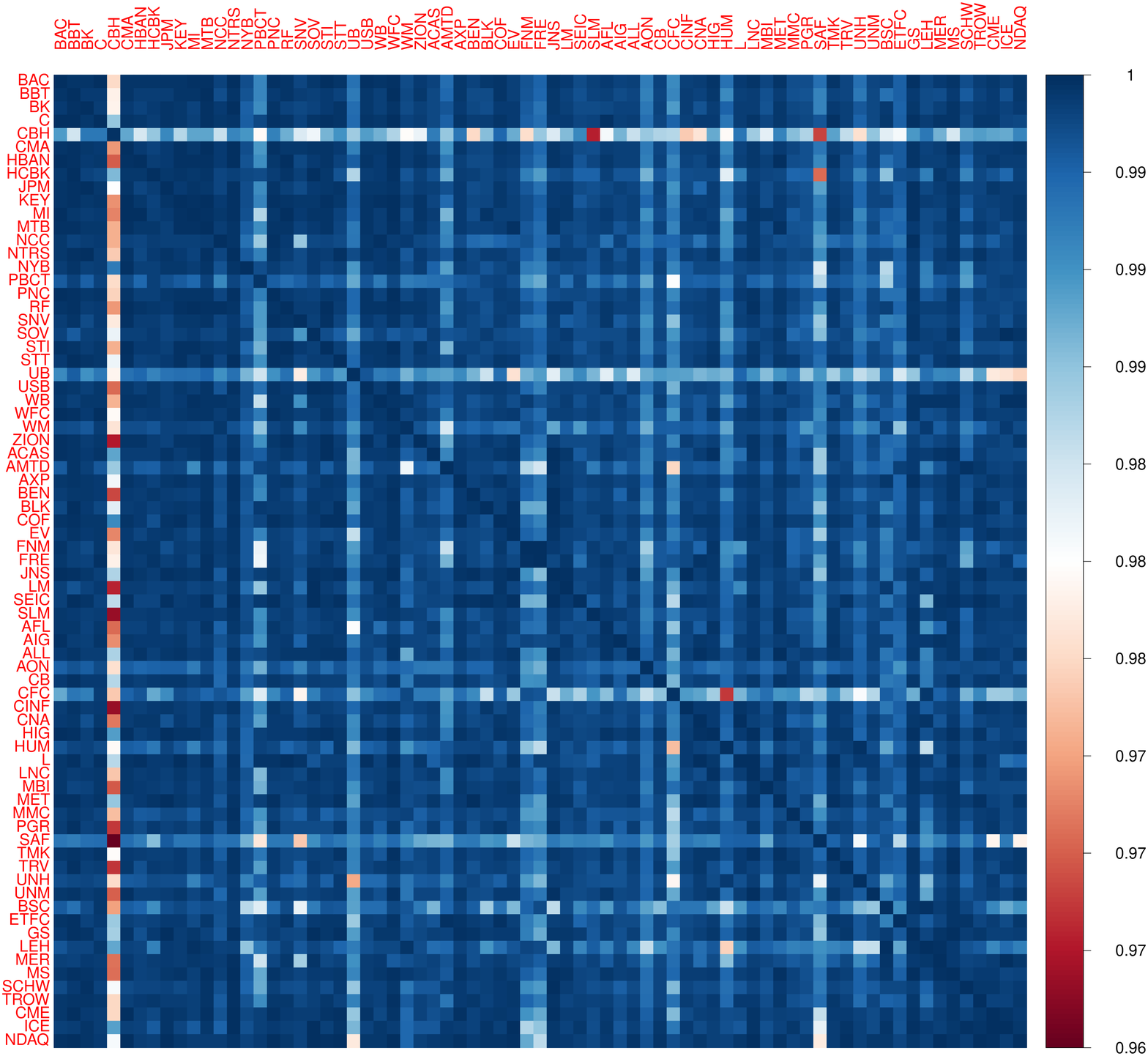}
    \caption{ Unconditional $\omega$ estimates of daily losses between the 73 institutions.}
    \label{fig:ommat.pdf}
\end{figure}

\begin{figure}[H]
     \centering
     \subfigure[]{
     \includegraphics[width=0.45\textwidth]{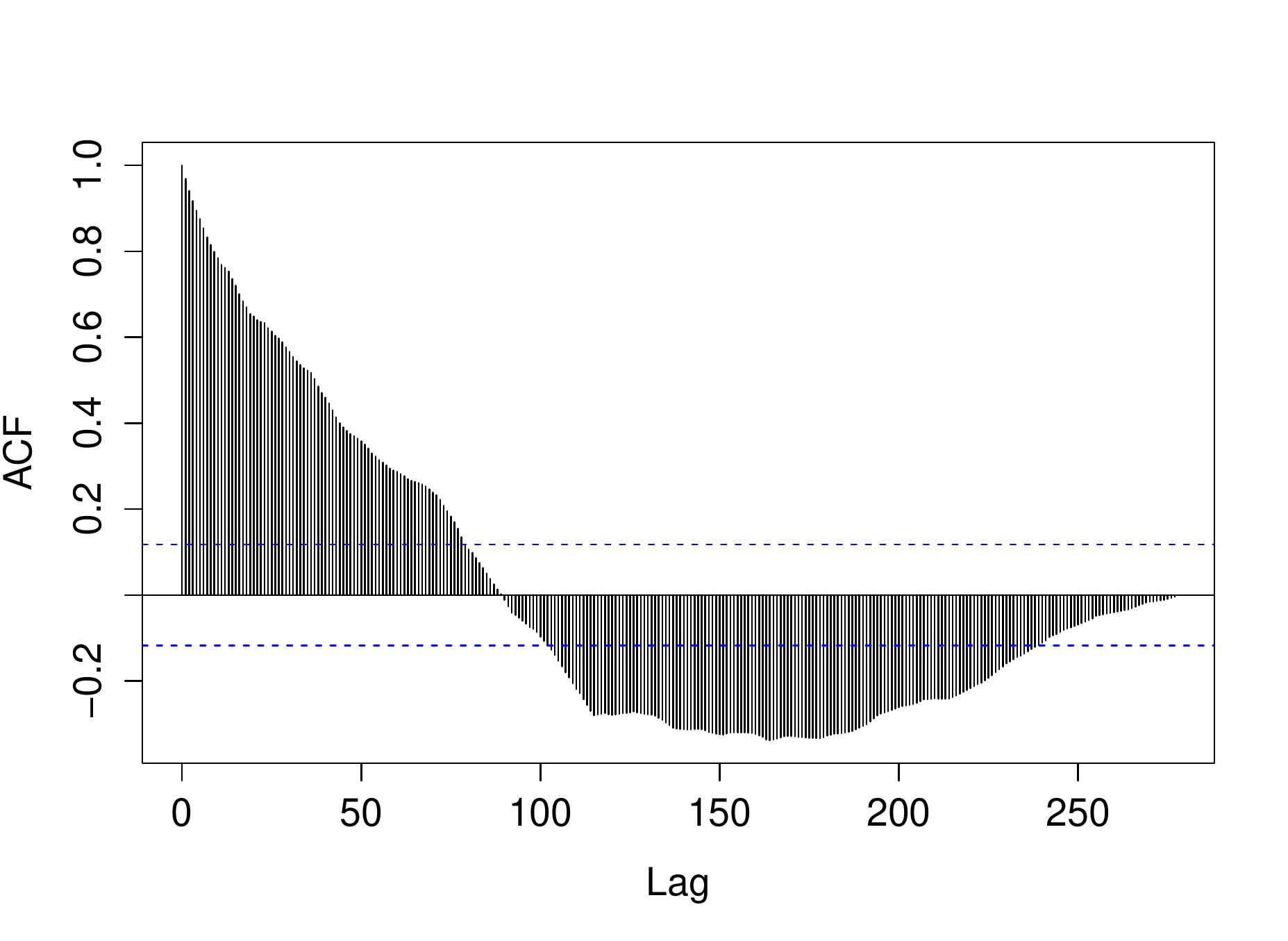}
     \label{fig:jpmdcov}
     }
     \qquad
        \subfigure[]{
     \includegraphics[width=0.45\textwidth]{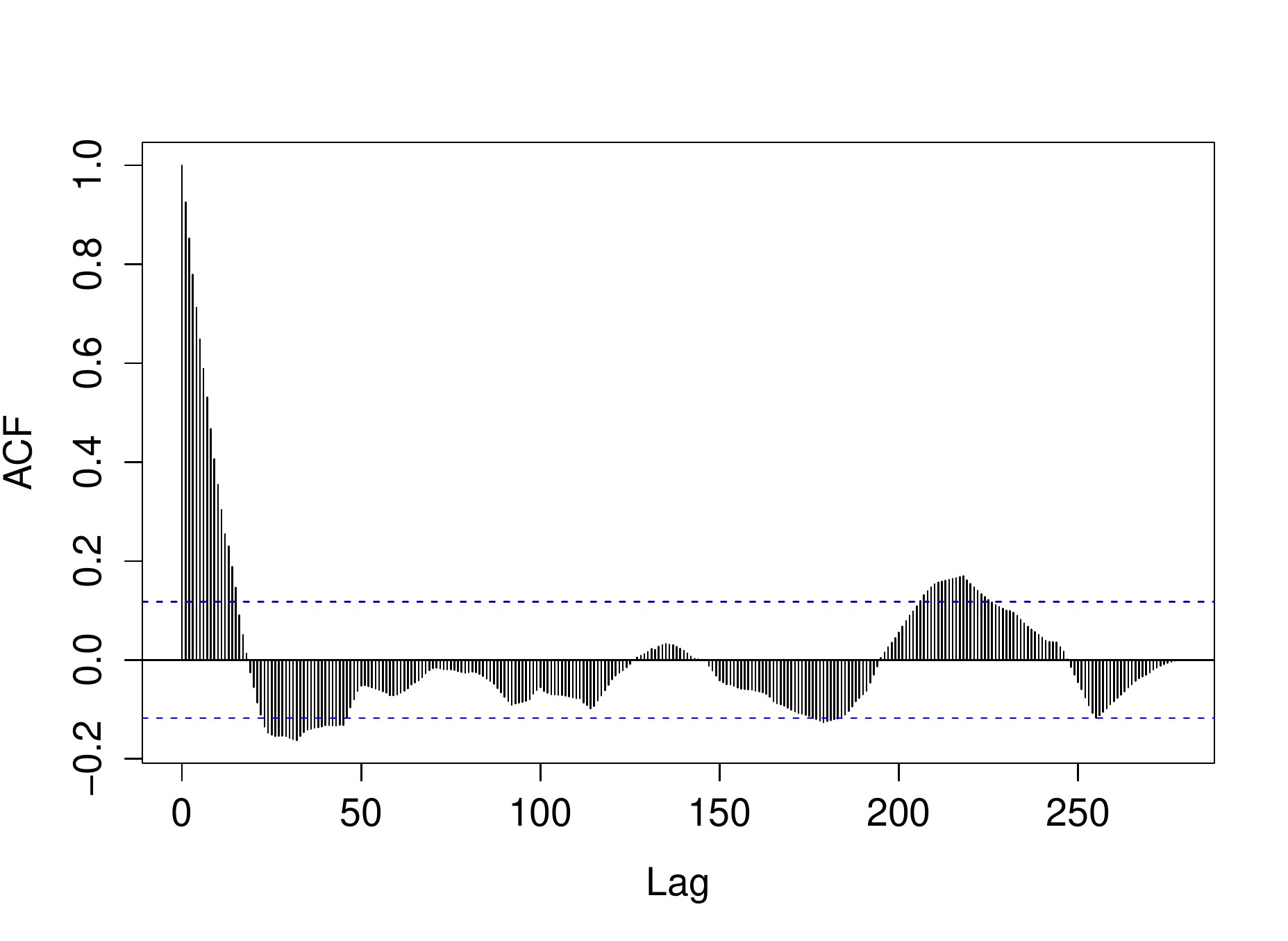}
     \label{fig:jpmdcoes}
     }
     \qquad
     \subfigure[]{
     \includegraphics[width=0.45\textwidth]{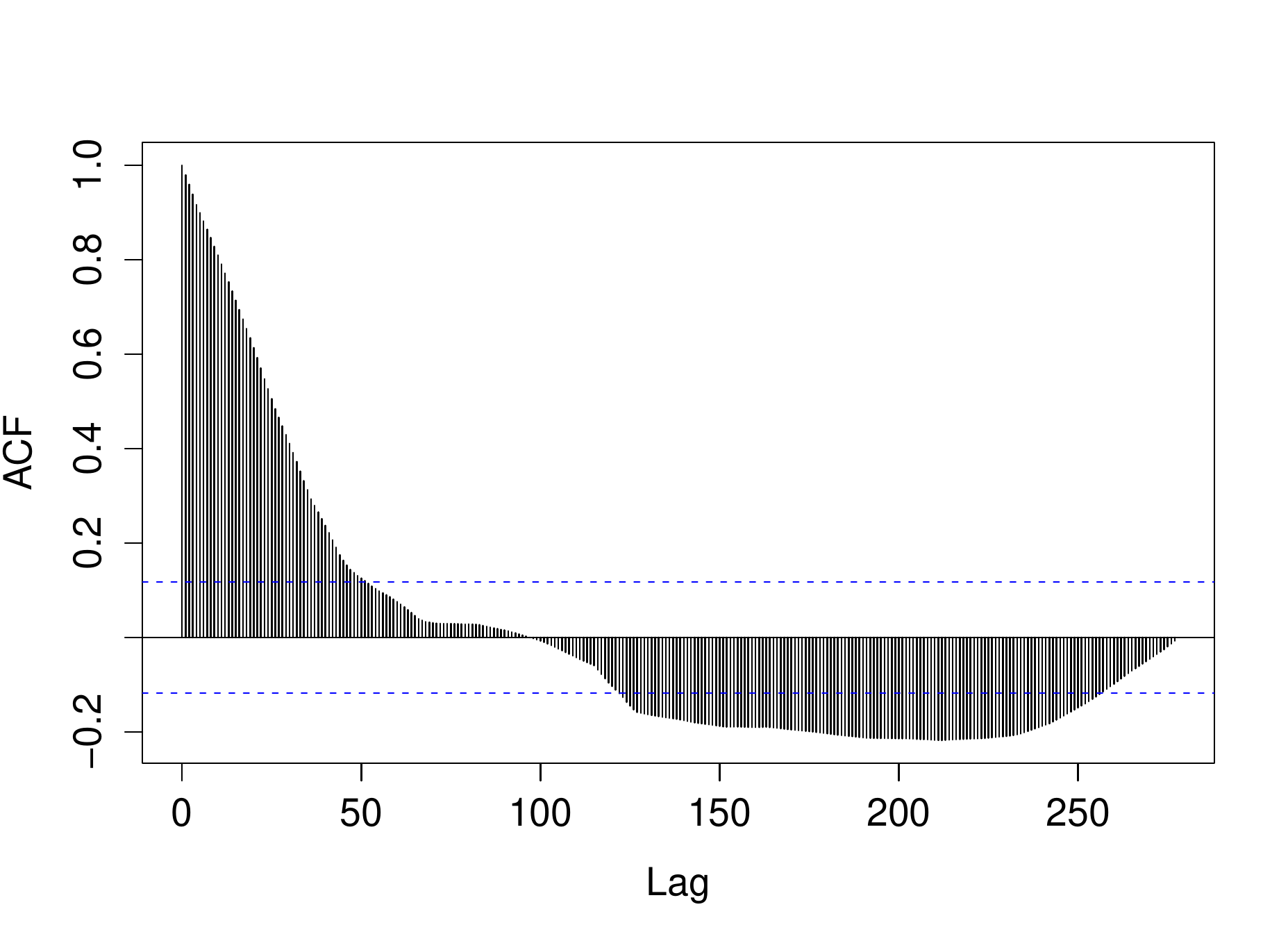}
     \label{fig:lehdcov}

     }
     \qquad
     \subfigure[]{
     \includegraphics[width=0.45\textwidth]{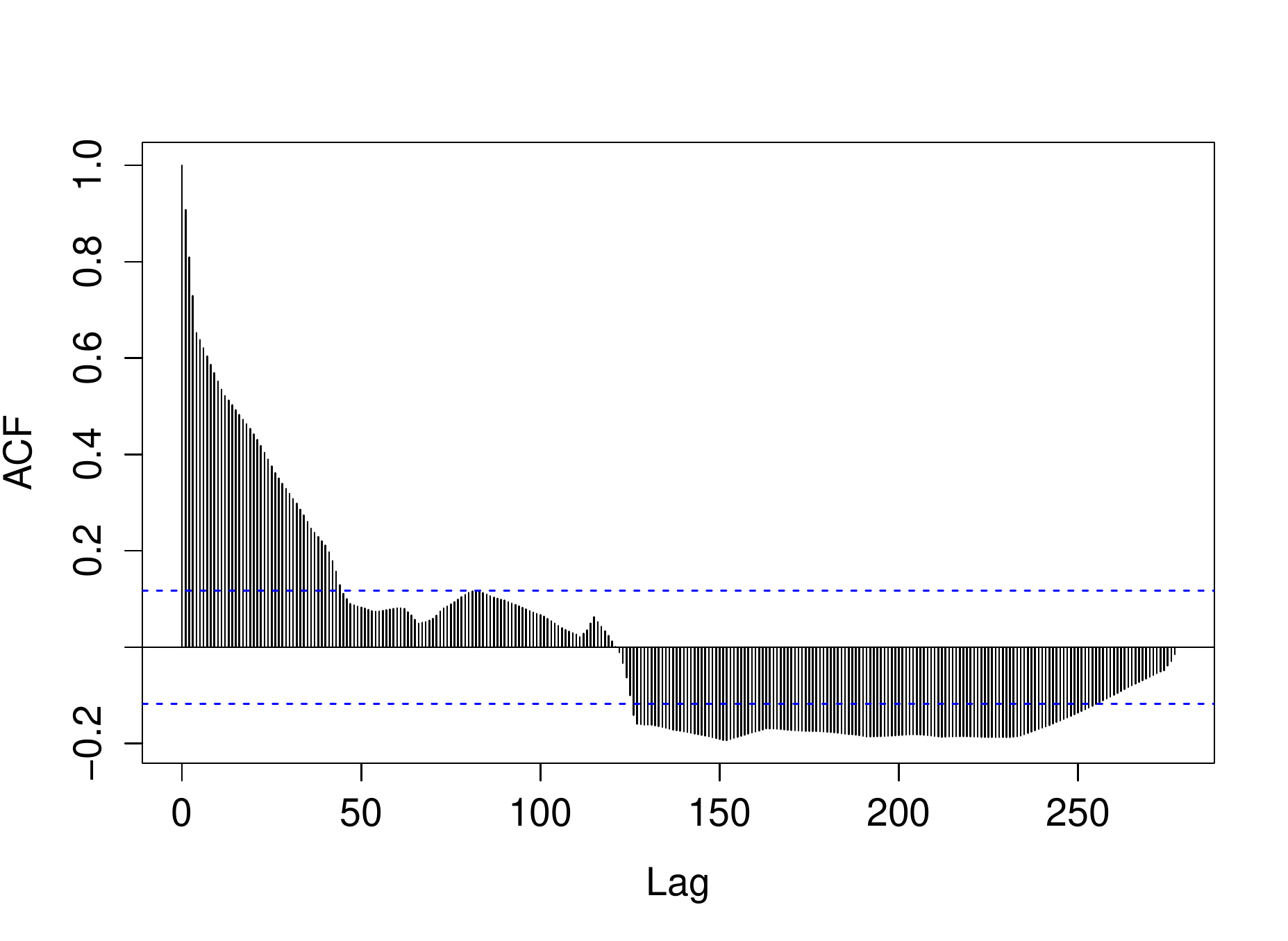}
     \label{fig:lehdcoes}
     }
     
     \caption{ACF plots of the average $\DCov$ of JP Morgan Chase \ref{fig:jpmdcov}, the average $\DCoES$ of JP Morgan Chase \ref{fig:jpmdcoes},the average $\DCov$ of Lehman Brothers \ref{fig:lehdcov} and the average $\DCoES$ of Lehman Brothers \ref{fig:lehdcoes}. Note the quicker decay of the autocorrelations of the $\DCoES$.}
     \label{fig:acfs}
 \end{figure}
\subsection{Details of computer and R setup}\label{App:compR}
Computer setup: 
\begin{itemize}
    \item HP Elitebook 2020
    \item CPU: AMD Ryzen 7 PRO 4750U with Radeon Graphics  @ 1.70 GHz
    \item RAM: 32 GB
    \item OS: Windows 10 Enterprise 21H2 build: 19044.1466
\end{itemize}
Software setup:
\begin{itemize}
    \item RStudio 2022.02.2+485 "Prairie Trillium" Release (8acbd38b0d4ca3c86c570cf4112a8180c48cc6fb, 2022-04-19) for Windows
Mozilla/5.0 (Windows NT 10.0; Win64; x64) AppleWebKit/537.36 (KHTML, like Gecko) QtWebEngine/5.12.8 Chrome/69.0.3497.128 Safari/537.36
\item R version: 4.2.0 (22-04-2022)\cite{Rsoftware}
\end{itemize}
R packages:
\begin{itemize}
    \item copula \cite{Coppackmanual}
    \item parallel (part of base R) \cite{Rsoftware}
    \item foreach \cite{foreach}
    \item xts \cite{xts}
    \item zoo \cite{Rzoo}
    \item tidyverse \cite{RTidyverse}
    \item dplyr \cite{Rdplyr}
    \item xtable \cite{Rxtable}
    \item haven \cite{Rhaven}
    \item corrplot \cite{Rcorrplot2021}
\end{itemize}

\end{document}